\documentclass[a4paper,12pt]{article}
\pdfoutput=1
\usepackage{graphicx, rotating,amssymb}

\ifx\pdfoutput\undefined
\usepackage[dvips,bookmarks=false]{hyperref}	
\else
\usepackage{hyperref}	
\fi
\hypersetup{colorlinks,bookmarksopen,bookmarksnumbered,citecolor=verdes,
linkcolor=blus,pdfstartview=FitH,urlcolor=rossos}
\def\myurl#1#2{\href{http://#1}{#2}}
\def\hhref#1{\href{http://arxiv.org/abs/#1}{#1}} 

\usepackage{multicol}
\usepackage{color}
\definecolor{rosso}{cmyk}{0,1,1,0.4}
\definecolor{rossos}{cmyk}{0,1,1,0.55}
\definecolor{rossoc}{cmyk}{0,1,1,0.2}
\definecolor{blu}{cmyk}{1,1,0,0.3}
\definecolor{blus}{cmyk}{1,1,0,0.6}
\definecolor{bluc}{cmyk}{1,1,0,0.1}
\definecolor{verde}{cmyk}{0.92,0,0.59,0.25}
\definecolor{verdec}{cmyk}{0.92,0,0.59,0.15}
\definecolor{verdes}{cmyk}{0.92,0,0.59,0.4}
\def\lsim{\raise0.3ex\hbox{$\;<$\kern-0.75em\raise-1.1ex\hbox{$\sim\;$}}}
\def\gsim{\raise0.3ex\hbox{$\;>$\kern-0.75em\raise-1.1ex\hbox{$\sim\;$}}}

\font\tenrsfs=rsfs10 at 12pt
\font\sevenrsfs=rsfs7
\font\fiversfs=rsfs5
\newfam\rsfsfam
\textfont\rsfsfam=\tenrsfs
\scriptfont\rsfsfam=\sevenrsfs
\scriptscriptfont\rsfsfam=\fiversfs
\def\mathscr#1{{\fam\rsfsfam\relax#1}}
\def\FERMI{{\sf Fermi}}
\def\PAMELA{{\sf PAMELA}}
\def\HESS{{\sf HESS}}

\newcommand{\fig}[1]{~\ref{fig:#1}}
\def\d{{\rm d}}

\def\circa#1{\,\raise.3ex\hbox{$#1$\kern-.75em\lower1ex\hbox{$\sim$}}\,}

\newcommand{\sigv}{\langle \sigma v \rangle}

\newcommand{\beq}{\begin{equation}}
\newcommand{\eeq}{\end{equation}}

\def\circa#1{\,\raise.3ex\hbox{$#1$\kern-.75em\lower1ex\hbox{$\sim$}}\,}
\makeatletter

\def\art{\@ifnextchar[{\eart}{\oart}}
\def\eart[#1]#2#3#4#5#6{{\rm #2}, {#3 #4} {\rm (#6) #5} [{\hhref{#1}}]}
\def\hepart[#1]#2{{\rm #2, \hhref{#1}}}
\newcommand{\oart}[5]{{\rm #1}, {#2 #3} {\rm (#5) #4}}

\newcounter{alphaequation}[equation]
\def\thealphaequation{\theequation\hbox to
0.6em{\hfil\alph{alphaequation}\hfil}}
\def\eqnsystem#1{
\def\@eqnnum{{\rm (\thealphaequation)}}
\def\@@eqncr{\let\@tempa\relax \ifcase\@eqcnt \def\@tempa{& & &} \or
  \def\@tempa{& &}\or \def\@tempa{&}\fi\@tempa
  \if@eqnsw\@eqnnum\refstepcounter{alphaequation}\fi
\global\@eqnswtrue\global\@eqcnt=0\cr}
\refstepcounter{equation} \let\@currentlabel\theequation \def\@tempb{#1}
\ifx\@tempb\empty\else\label{#1}\fi
\refstepcounter{alphaequation}
\let\@currentlabel\thealphaequation
\global\@eqnswtrue\global\@eqcnt=0 \tabskip\@centering\let\\=\@eqncr
$$\halign to \displaywidth\bgroup \@eqnsel\hskip\@centering
$\displaystyle\tabskip\z@{##}$&\global\@eqcnt\@ne
\hskip2\arraycolsep\hfil${##}$\hfil& \global\@eqcnt\tw@\hskip2\arraycolsep
$\displaystyle\tabskip\z@{##}$\hfil
\tabskip\@centering&\llap{##}\tabskip\z@\cr}
\def\endeqnsystem{\@@eqncr\egroup$$\global\@ignoretrue} \makeatother

\begin{document}
\begin{center}
{CERN-PH-TH/2009-224}
{ \hfill SACLAY--T09/187}
\color{black}
\vspace{0.5cm}

{\LARGE \bf Diffuse gamma ray constraints on annihilating\\[2mm] or decaying Dark Matter after Fermi}

\bigskip\color{black}\vspace{0.6cm}

{
{\large\bf Marco Cirelli}$^{a,b}$,
{\large\bf Paolo Panci}$^{c,d}$,
{\large\bf Pasquale D. Serpico}$^{a,e}$
}
\\[5mm]
{\it $^a$ CERN Theory Division, CERN, \\ 
Case C01600, CH-1211 Gen\`eve, Switzerland}\\[3mm]	
{\it $^b$ Institut de Physique Th\'eorique, CNRS, URA 2306 \& CEA/Saclay, \\
	F-91191 Gif-sur-Yvette, France}\\[3mm]
{\it $^c$ Dipartimento di Fisica, Universit\`a degli Studi dell' Aquila, 67010 Coppito (Aq), and\\
INFN, Laboratori Nazionali del Gran Sasso, 67010 Assergi (Aq), Italy}\\[3mm]
{\it $^d$ Universit\'e Paris 7-Diderot, UFR de Physique,\\ B\^atiment Condorcet, 10, rue A. Domon et L.Duquet, 75205 Paris, France}\\[3mm]
{\it $^e$ LAPTH, UMR 5108 CNRS,\\ 9 chemin de Bellevue - BP 110, 74941 Annecy-Le-Vieux, France}
\end{center}

\medskip

\centerline{\large\bf Abstract}
\vspace{-0.3cm}
\begin{quote}
\color{black}\large
We consider the diffuse gamma ray data from \FERMI\ first year observations and compare them to the gamma ray fluxes predicted by Dark Matter annihilation or decay (both from prompt emission and from Inverse Compton Scattering), for different observation regions of the sky and a range of Dark Matter masses, annihilation/decay channels and Dark Matter galactic profiles. 
We find that the data exclude large regions of the Dark Matter parameter space not constrained otherwise and
discuss possible directions for future improvements. 
Also, we further constrain Dark Matter interpretations of the $e^\pm$ \PAMELA/\FERMI\ spectral anomalies, both for the annihilating and the decaying Dark Matter case: under very conservative assumptions, only  models producing dominantly $\mu^{\pm}$ and assuming a cored Dark Matter galactic profile can fit the lepton data with masses around $\sim 2\,$TeV.
\end{quote}


\vspace{-0.3cm}

\section{Introduction}
\label{introduction}
A wealth of astrophysical and cosmological data have revealed the crucial gravitational role of otherwise essentially non-interacting particles, dubbed Dark Matter (DM). This provides one of the most tantalizing hints for physics beyond the Standard Model (SM). However, in order to identify the nature of these particles, one needs information on their mass and interaction properties. One strategy is to explore the possible indirect signatures coming from their Standard Model annihilation (or decay) products in the halo of our Galaxy, or beyond. For typical candidates in the category of weakly interacting massive particles (WIMPs) these products fall in the realm of investigation of high-energy astrophysics (typically $E\gsim 1\,$GeV), as gamma-rays, neutrinos, etc.

\medskip

The Fermi Gamma-ray Space Telescope~\cite{FERMIurl} team has recently released its first year data. While the analyses for their detailed  interpretation in terms of known astrophysical sources are still ongoing, preliminary spectra from different regions of the sky have been presented at several conferences, most notably the \FERMI\ Symposium~\cite{symp}. These spectra extend to comparable or higher energies than the predecessor mission {\sf EGRET}, but above all have significantly reduced systematics. In particular, ``anomalous'' (with respect to simple propagation models predictions) bumps of radiation above the  GeV have not been confirmed, see e.g.~\cite{noexcess}. 

\medskip

Given these new data, it is timely to provide a first assessment of their power  in constraining DM properties. Rather than considering a specific scenario beyond the Standard Model, we adopt a phenomenological approach, namely: i) assume DM annihilation or decay in a single dominant channel ($e^+e^-$, $\mu^+\mu^-$, $\tau^+\tau^-$, $W^+W^-$, $b\bar{b}$ or $t\bar{t}$); ii) for different choices of the galactic DM halo profile, derive constraints in the $m_\chi$--$\sigv$ or $m_\chi$--$\tau_{\rm dec}$ plane, for the cases of self-annihilating or decaying Dark Matter respectively. Here $m_\chi$ is the mass of the DM particle $\chi$, $\sigv$ its self-annihilation cross section and $\tau_{\rm dec}$ its half life (for the decay case, we limit to the leptonic channels, as discussed below). 

In Sec.~\ref{data} we deal with a short description of the formalism, the datasets used and the choices for the galactic Dark Matter halo profile, while the constraints that we obtain are reported in Sec.~\ref{results}.  
In deriving our bounds we assume no prior information from other cosmic ray data, nor we worry about astrophysical sources of gamma rays which are surely present and probably even dominating the signal. This is clearly an overconservative approach: in Sec.~\ref{discussion} we comment  on the expected reach in exploring the DM parameter space after a better understanding of the astrophysical background is achieved. 

\smallskip

Since, motivated by peculiar spectral features, a plethora of models have appeared trying to fit  the positron fraction and total electron data from \PAMELA, \FERMI\ and \HESS~\cite{PAMELApositrons, FERMIleptons, HESSleptons, HESSleptons2} in terms of Dark Matter (as we review in Sec.~\ref{PAMFERMI}), in Sec.~\ref{results} we also comment upon the consistency of these scenarios in the light of the new diffuse gamma ray data, as a specific application of the general constraints derived above. 
We limit ourselves to modes of annihilation in a pair of charged lepton final states at tree level since, on one hand, they appear the only ones not excluded by other constraints (most notably antiproton fraction~\cite{CKRS, Salatipbar}), on the other hand they are significantly less model-dependent than scenarios with annihilations proceeding via additional, non SM states~\cite{onestep}. 
For the case of {\it decaying} DM, which has been invoked as alternative explanation of the lepton anomalies, a fortiori we limit ourselves to leptonic channels.   We anticipate that explanations based on DM models are mostly excluded as leading contributors to the \PAMELA / \FERMI\ features. The only exceptions (but see discussion in  Sec.~\ref{discussion}) appear to be for models assuming mostly $\mu^{\pm}$ final states and, for annihilating models, also requires cored galactic DM profiles.
Note that even models producing dominantly $\tau^{\pm}$, previously allowed, are now excluded (or strongly disfavoured for sufficiently cored profiles) by the data: this applies also to decaying scenarios.

A discussion of the caveats and further constraints, perspectives, etc. is reported in Sec.~\ref{discussion}, together with our conclusions.

\vspace{-0.1cm}

\section{Datasets and constraints} 
\label{data}

\vspace{-0.2cm}

\subsection{Calculation of the signal}
The differential flux of photons from (self-conjugated) dark matter annihilations in the galactic halo, observed from a given direction in the sky, is written as
\begin{equation}
\frac{\d \Phi_{\rm halo}^{\rm ann}}{\d \epsilon \ \d \Omega}=
\,\frac{1}{2} \frac{\sigv}{4\,\pi} \, r_\odot \frac{\rho_\odot^2}{m_\chi^2} \int_{\rm
los} \d s\, \frac{1}{r_\odot}\,\left(\frac{\rho_{\rm halo}[r(s,\psi)]}{\rho_\odot}\right)^2\,\frac{\d N}{\d \epsilon}\,, \label{Ism}
\end{equation}
where the coordinate $r$, centered on the Galactic Center, reads $r(s,\psi)=(r_\odot^2+s^2-2\,r_\odot\,s\cos\psi)^{1/2}$,
$r_\odot = 8.33$ kpc~\cite{Gillessen:2008qv}
 is the most likely distance from the Sun to the Galactic Center (GC) and
$\psi$ is the angle between the direction of observation in the sky and the GC. In terms of the galactic latitude $b$ and longitude $l$, one has
$
\cos\psi=\cos b\cos l\,.
$
The coordinate $s$ parameterizes the distance from the Sun along the line-of-sight (los). 
The gamma ray energy is denoted by $\epsilon$.
Particle physics enters via the DM mass $m_\chi$, the annihilation cross section $\sigv$, and
the photon differential energy spectrum $\d N/\d \epsilon$
per annihilation. 

This spectrum is in general constituted of two main components: i) the prompt gamma rays, originating from the fragmentation of the primary products of annihilation (essentially via bremsstrahlung of charged particles in the shower and production of $\pi^0$ and their subsequent decay into $\gamma\gamma$), and ii) the Inverse Compton (IC) gamma rays, produced by the upscattering of the low energy photons of the starlight, the infrared light and the CMB (denoted collectively as the InterStellar Radiation Field, ISRF) by the energetic $e^\pm$ injected by Dark Matter annihilations.\footnote{Other contributions, such as a line component from loop-suppressed annihilations into $\gamma\gamma, \gamma Z$ or $\gamma h$ (with $Z$ and $h$ the Standard Model gauge and Higgs boson) and Internal Bremsstrahlung gamma rays from the charged particles in the annihilation diagrams could be present, but, as they are model dependent, we do not consider them here. In general they introduce features in the total gamma ray spectrum that would make stronger the constraints from an analysis like ours.} We compute the IC gamma ray spectrum as discussed in detail in~\cite{CP}, adopting in particular the approximation of neglecting the diffusion processes of the $e^\pm$ injected by Dark Matter. While this can have an impact on small internal galactic regions (where the DM density has the highest gradient), this is instead a very good approximation for large observational regions away from the GC (amounting to a correction of less than a factor of 2 on the amplitude of the flux already for intermediate latitudes~\cite{CP,MPSV}). 
The total spectrum $\d N/\d \epsilon$ is strictly independent of direction only for the prompt component, while for the IC it carries a dependence on the characteristics of the background light on which the IC scattering occurs.
We will however work under the approximation of constant ISRF over each of the observation regions that we consider (see below), as already done in~\cite{CP}; whenever the region is away from the GC, the dominant background for IC processes is in any case the Cosmic Microwave Background (CMB), which is obviously identical in each region. The flux from a region $\Delta \Omega$ of the sky therefore simply rearranges in terms of the usual average geometrical factor $\bar J_{\rm ann}$ as
\begin{equation}
\frac{\d \Phi_{\rm halo}^{\rm ann}}{\d \epsilon}=
\,\frac{1}{2} \frac{\sigv}{4\,\pi} \, r_\odot \frac{\rho_\odot^2}{m_\chi^2}\,  \bar J_{\rm ann} \Delta \Omega\, \frac{\d N}{\d \epsilon}, \qquad
\bar J_{\rm ann} \Delta \Omega = \int_{\Delta \Omega} \d \Omega(b,l) \int_{\rm los} \frac{\d s}{r_\odot}\,\left(\frac{\rho_{\rm halo}[r(s,\psi)]}{\rho_\odot}\right)^2\,.
\end{equation}

\medskip

In the case of decaying dark matter one has 
\begin{equation}
\frac{\d \Phi_{\rm halo}^{\rm dec}}{\d \epsilon \ \d \Omega} = \frac{\Gamma}{4\pi} r_\odot \frac{\rho_\odot}{m_\chi} \int_{\rm{los}} \d s \frac{1}{r_\odot} \left( \frac{\rho_{\rm halo}[r(s,\psi)]}{\rho_\odot} \right) \frac{\d N}{\d \epsilon}\,,
\label{fluxdec}
\end{equation}
and thus (the same discussions as above apply)
\begin{equation}
\frac{\d \Phi_{\rm halo}^{\rm dec}}{\d \epsilon} = \frac{\Gamma}{4\pi} r_\odot \frac{\rho_\odot}{m_\chi}   \bar J_{\rm dec} \Delta \Omega \, \frac{\d N}{\d \epsilon}\,, \qquad \bar J_{\rm dec} \Delta \Omega =  \int_{\Delta \Omega} \d \Omega(b,l)  \int_{\rm los}  \frac{\d s}{r_\odot} \left( \frac{\rho_{\rm halo}[r(s,\psi)]}{\rho_\odot} \right) ,
\label{fluxdecint}
\end{equation}
where $\Gamma=\tau_{\chi}^{-1}$ is the decay rate (inverse lifetime) and the spectrum now refers to the photons generated in each decay process, including  both the prompt and the Inverse Compton emission.  

\medskip

For the decaying DM scenario, we will also need the cosmological flux, i.e. the flux of gamma rays due to cosmological DM decays integrated over redshift. As we will see, it leads to powerful constraint and ---differently from the analogous annihilating DM signal---it does not depend on halo profiles or DM substructures, relying only on robustly known quantities. It is straightforward to show that the differential spectrum in Earth-measured energy $\epsilon$ from the entire sky writes
\begin{equation}
\frac{\d \Phi^{\rm dec}_{\rm cosm}}{\d \epsilon}=\Gamma\, \frac{\Omega_{\rm DM}\,\rho_{c,0}}{m_\chi} 
\,\int_0^\infty \d z\,\frac{e^{-\tau(\epsilon(z),z)}}{H(z)} \frac{\d N}{\d \epsilon}(\epsilon(z),z)\,,\label{smoothedmap}
\end{equation}
where the Hubble function, $H(z)$, writes explicitly $H(z)=
H_0\sqrt{\Omega_M(1+z)^3+\Omega_\Lambda}$ where $H_0$ is  the present Hubble
expansion rate and $\Omega_M$ and  $\Omega_\Lambda$ are respectively the matter and 
cosmological constant energy density in units of the critical density, $\rho_{c,0}$.
This flux originates from the integral along the time of travel of a photon emitted in the past, when converted as an integral in redshift $z$ according to the standard relation $\d t/\d z = 1/((1+z)\,H(z))$; the $(1+z)$ factor is canceled by an identical $(1+z)$ at the numerator due to the redshift in the energy of the same photon. 
In turn, the rescaling factor with redshift of the DM density ($(1+z)^3$) cancels with an identical factor coming from the volume expansion of the Universe. Finally, the factor $e^{-\tau(\epsilon,z)}$ accounts for the finite optical depth, $\tau$, of
the universe to high energy gamma rays due to scattering with the extragalactic background light (see e.g.~\cite{CIP} for a more detailed discussion). For energies $\epsilon \lsim 100\,$GeV, the latter has only a moderate impact.
The gamma ray spectrum $\d N/\d \epsilon$ is again the sum of the prompt (${\rm P}$) and IC contributions, as produced at any redshift $z$. The dependence on $z$ in the prompt component is just due to the redshifting of the energy in the argument, i.e. $\d N^{\rm P}/\d \epsilon\, (\epsilon(z),z)=\d N^{\rm P}/\d \epsilon\, (\epsilon(1+z),0)$. For the IC component, the ISRF consists in this case of CMB photons only, rescaled in energy according to the redshift. The integral over redshift is dominated by the range up to $z \approx 20$.  We notice that, in the (well fulfilled) limit where the IC loss time is much faster than the Hubble time, the electrons lose all their energy via IC, and all scatterings happen in the Thomson regime,  one has
\begin{equation}
 \frac{\d N^{\rm IC}}{\d \epsilon}(\epsilon(z),z)\simeq \frac{1}{1+z} \frac{\d N^{\rm IC}}{\d \epsilon}(\epsilon,0)\,.
\end{equation}
The above formula can be also understood heuristically as follows: the IC cross section in the Thomson regime is energy-independent; since the energy spectrum of the injected particles is the same at any $z$, the IC photon at the production epoch $z$ will have an energy proportional to the one of the upscattered background photon.
The latter is $(1+z)$ times the current one, which is exactly the factor compensated by the subsequent redshifting. So, the IC spectrum is universal and equal to the one calculated at $z=0$. The remaining factor $(1+z)^{-1}$ can be immediately understood since $\d t=\d z/(H(z)(1+z))$: apart for the absorption correction, the integral in Eq.~(\ref{smoothedmap}) for the IC is the  ``spectrum per decay event" times the number density $\Omega_{\rm DM}\,\rho_{c,0}/m_\chi$ of the metastable particles, times the probability of decay over the lifetime of the universe $t_U$, i.e. $t_U\,\Gamma$.

The \FERMI\ collaboration has presented preliminary results on
the (maximal) residual, isotropic gamma-ray flux present in their data (see below). When using these diffuse data to constrain decaying DM properties, one has to account for the fact that the DM decays contribute with two terms to the above flux: i) a truly  cosmological flux, which is of course isotropic at leading order. ii) the residual  emission at high latitudes from the DM halo of our Galaxy: for the approximation of spherical halo, this is minimum in the anti-GC and grows very slowly towards 
the intermediate latitudes (see e.g.~\cite{SerpicoMiniReview}). In formul\ae, the predicted DM flux that we will compare with \FERMI\ isotropic diffuse $\gamma$-ray data is
\begin{equation}  
\frac{\d \Phi^{\rm dec}_{\rm isotropic}}{\d \epsilon} = \frac{\d \Phi^{\rm dec}_{\rm cosm}}{\d \epsilon} + 4 \pi \left. \frac{\d \Phi^{\rm dec}_{\rm halo}}{\d \epsilon\, \d \Omega}\right|_{\rm anti-GC}
\end{equation}  
For typical DM decay channels and for any DM halo profile, we find that the two contributions are of comparable amplitude.
Note that a similar equation would hold for the annihilating DM case with two significant differences:
a) the halo signal has a stronger dependence from the angular distance from the GC; b) while for the decaying case the cosmological signal is readily calculated in terms of known quantities, see Eq.~(\ref{smoothedmap}), for the annihilating case a dependence is introduced on the halo profiles and the clumpiness of DM halos: this is the reason why we do not include this constraint in the following, but in Sec.~\ref{discussion} we shall discuss its likely importance for cored halos. For a  comparison
of the role of the two terms in the annihilating case, see also~\cite{Serpico:2007}.
\bigskip

For the galactic distribution of Dark Matter we consider the cases of a Navarro, Frenk and White (NFW)~\cite{Navarro:1995iw}, Einasto~\cite{Graham:2005xx, Navarro:2008kc} and cored Isothermal~\cite{Bahcall:1980fb} profiles. The first (peaked as $r^{-1}$ at the GC) is a traditional benchmark choice motivated by N-body simulations, the second (not converging to a power law at the GC and more chubby than NFW at the location of the Sun) is emerging as a better fit to more recent numerical simulations. Cored profiles, such as the Isothermal one or the Burkert profile~\cite{Burkert}, might be instead more motivated by the observations of galactic rotation curves, but seem to run into conflict with the results of numerical simulations. As long as a convergent determination of the actual DM profile is not reached, it is worth considering a range of possible choices. The functional forms of these profiles read
\begin{equation}
 \rho_{\rm NFW}(r)=\rho_{s}\frac{r_{s}}{r}\left(1+\frac{r}{r_{s}}\right)^{-2},
   \label{eq:NFW}
\end{equation}

\begin{equation}
 \rho_{\rm Ein}(r)=\rho_{s}\exp\left\{-\frac{2}{0.17}\left[\left(\frac{r}{r_{s}}\right)^{0.17}-1\right]\right\},
   \label{eq:Einasto}
\end{equation}

\begin{equation} 
   \rho_{\rm isoT}(r)=\frac{\rho_{s}}{1+\left(r/r_{s}\right)^{2}}. 
   \label{eq:isoT}
\end{equation}

A separate but related issue is the one of fixing the parameters $r_s$ and $\rho_s$ that enter in each of these forms. We apply the following criterion. We fix the NFW parameters at the benchmark values $r_s = 20$ kpc and $\rho_s = 0.26$. This produces a DM halo with a density of $\rho_\odot = 0.3$ GeV/cm$^{3}$ and with a total DM mass contained in 50 kpc (i.e. within the distance of the Large Magellan Cloud) of $M_{50}\equiv 3.6\times \, 10^{11} M_\odot$. We then impose that the other profiles obey the same constraints on $\rho_\odot$ and $M_{50}$. This is motivated by the following considerations: on one hand, the total mass estimate within 50 kpc is quite robust~\cite{Sakamoto:2002zr}, on the other hand  recent analyses suggest that the best-fit value for the density of the DM at the solar distance is the same for different models to within $\sim 7\%$~\cite{Catena:2009mf}. The parameters that we adopt are thus given explicitly by:
$$ \begin{tabular}{l|cc}
  DM halo model & $r_{s}$ in kpc & $\rho_{s}$ in GeV/cm$^{3}$\\
  \hline
  NFW & 20 & 0.26\\
  Einasto & 21.8 & 0.05\\
  Isothermal & 3.2 & 2.31
 \end{tabular}$$
 We emphasize that in no way we are claiming that the choices for the parameters $\{\rho_s,M_{50}\}$
 are the optimal ones, but by the above criteria we are ``normalizing''  the alternative profiles in a more physically motivated way, commensurate with the benchmark profile which is often employed in the literature. In any case, we note that employing slightly different normalizations (such as those already employed in~\cite{CP}) has an overall very small impact on the analysis.

\subsection{\FERMI\ diffuse gamma-ray observations}
\label{FERMIdatasets}

\begin{table}[t]
\centering
\footnotesize{
\begin{tabular}{l|c|ccc|ccc|c}
Region & \multicolumn{1}{c|}{latitude $b$ \&}  &  \multicolumn{3}{c}{$\bar J_{\rm ann} $} & \multicolumn{3}{|c|}{$\bar J_{\rm dec}$} & ISRF\\
 & \multicolumn{1}{c|}{longitude $l$} &  IsoT & NFW  & Einasto & IsoT & NFW  & Einasto & \\
\hline
\vspace{-3mm} & & & & & & & \\
`3$\times$3'   &$0^\circ < $ $|b|$ $< 3^\circ$  & 35 & 374 & 635 & $-$ & $-$ & $-$ & GC\\
 & $357^\circ < $ $ l$   $< 360^\circ$ &  & & & & & \\
  & $0^\circ < $ $ l $   $< 3^\circ$ & & & & & & \\[1mm]
\hline
\vspace{-3mm} & & & & & & & \\
`5$\times$30'   &$0^\circ < $ $|b|$ $< 5^\circ$  & 21 & 56 & 93 & $-$ & $-$ & $-$ & GC\\
 & $330^\circ < $ $ l$   $< 360^\circ$ &  & & & & & \\
  & $0^\circ < $ $ l $   $< 30^\circ$ & & & & & & \\[1mm]
\hline 
\vspace{-3mm} & & & & & & & \\
`10$-$20' & $10^\circ <$ $|b|$ $< 20^\circ$  &  3.35 & 3.46 & 4.23 & 2.45 & 2.38 & 2.47 & $h=5$ kpc \\
& $0^\circ < $ $l$ $ < 360^\circ$ & & & & & & \\
\hline
\vspace{-3mm} & & & & & & & \\
`Gal Poles' & $60^\circ <$ $|b|$ $< 90^\circ$  &  0.92 & 0.96 & 0.94 & 1.73 & 1.69 & 1.67 & CMB \\
& $0^\circ < $ $l$ $ < 360^\circ$ & & & & & & \\
\hline 
\end{tabular}
\caption{\em Summary of the observational regions that we consider, with the corresponding values of the average $\bar J$ factor for different DM halo profiles and the adopted Inter-Stellar Radiation Field.\label{tab:regions}}}
\end{table}

The \FERMI\ collaboration has presented results on diffuse gamma rays from the first year analysis in terms of data points in several different observational windows. For the annihilating Dark Matter case, we focus in particular on inner galaxy and intermediate latitude regions. Namely, we consider:

\begin{itemize}
\item[$\diamond$] A square region that includes the Galactic Center, at galactic latitude $0^\circ < |b| < 3^\circ$ and galactic longitude $357^\circ < l < 360^\circ$ and $0^\circ < l < 3^\circ$, that we denote as `$3^\circ \times 3^\circ$' region. We use the preliminary \FERMI\ data points as presented in~\cite{Digel}. 
\item[$\diamond$] A rectangular region that also includes the Galactic Center, at galactic latitude $0^\circ < |b| < 5^\circ$ and galactic longitude $330^\circ < l < 360^\circ$ and $0^\circ < l < 30^\circ$, denoted as `$5^\circ \times 30^\circ$' region~\cite{Digel}.
\item[$\diamond$] The intermediate latitude region defined by $10^\circ < |b| < 20^\circ$ in galactic latitude, at all galactic longitudes (denoted in the following as `$10^\circ-20^\circ$ strips'). The released \FERMI\ data points (from~\cite{noexcess}) improve on the former release in~\cite{oldstrip,Porter}, confirming the results.
\item[$\diamond$] The high latitude `Galactic Poles' region, identified by $|b| > 60^\circ$ at all longitudes. We use the data points from~\cite{AckermannTeVPA}. 
\end{itemize}

For the decaying Dark Matter case we consider the two intermediate/high latitude galactic regions discussed above (`$10^\circ-20^\circ$ strips' and `Galactic Poles' ) and, in addition:

\begin{itemize}
\item[$\diamond$] The measurements of the isotropic diffuse flux presented in~\cite{Ackermann}
(see~\cite{Chen:2009uq} for a similar kind of analysis, limited to decaying DM models).
\end{itemize}

In all the regions, the preliminary \FERMI\ data points that we use extend to about 100 GeV. Ongoing work in the collaboration will likely extend this range to higher energies. We comment on the impact of this below. 
The uncertainties that we use are generally quoted to include statistical and systematic errors, but a more detailed analysis is in the making by the collaboration.

\medskip

In order to compute the ICS signal from DM annihilations or decays, we model the ISRF for each region as discussed in~\cite{CP}, in terms of renormalized black body spectra that reproduce the background light characteristic of the different areas.
For the inner galaxy regions (`$3^\circ \times 3^\circ$' and `$5^\circ \times 30^\circ$') we use the ISRF computed for the Galactic Center ($R=0, h=0$, in galactic cylindrical coordinates) in~\cite{ISRF}, that features an important contribution at small wavelengths of starlight photons, because of the higher density of stars in the galactic bulge. 
For the `$10^\circ-20^\circ$ strips' we use the ISRF computed for distances above the galactic plane, $h=5$\, kpc, in~\cite{ISRF2}.
For the Galactic Poles and of course for the cosmological flux we use the CMB only.

\subsection{Details on the DM interpretations of the $e^\pm$ anomalies}
\label{PAMFERMI} 

As mentioned in the Introduction, the anomalous \PAMELA, \FERMI\ and \HESS\ data have been interpreted in terms of DM annihilations~\cite{CKRS, MPSV} or decay~\cite{decay}. We recall here briefly the main features of the experimental data and of their DM interpretations, without entering in the details of any specific particle physics model.

The \PAMELA\ satellite has reported a rise of the positron fraction $e^+/(e^++e^-)$ above the expected declining astrophysical background from $\sim$10~GeV up to at least 100~GeV~\cite{PAMELApositrons}.
At the same time, the ratio $\bar p/p$ fluxes is consistent with  expectations based on astrophysical secondaries, up to the maximal probed energy of about 200 GeV~\cite{PAMELApbar, AdrianiCOSMO09}. 
\FERMI~\cite{FERMIleptons,Latronico} has reported a $e^++e^-$ spectrum suggestive
of an additional harder component on the top of a smooth astrophysical spectrum, the latter falling slightly more steeply than $E^{-3}$. The \HESS\ \v Cerenkov telescope, too, has published data~\cite{HESSleptons, HESSleptons2} in the range of energy from 600 GeV up to a few TeV, showing a power law spectrum in agreement with the one from \FERMI\ and eventually a steepening at energies of a few TeV.

We perform the fits to these data\footnote{We consider the \PAMELA\ data for positrons at energies larger than 10 GeV only, where the uncertainty due to solar modulation is not present. We use the new preliminary analysis of $\bar p$ data from \PAMELA, as presented e.g. in~\cite{AdrianiCOSMO09}, that extends to about $\sim$ 200 GeV and provides somewhat reduced uncertainty bars. We include the systematic error band on the \FERMI\ and \HESS\ datapoints. We add the low energy datapoints in the $e^++e^-$ from \FERMI\ as presented in~\cite{Latronico}: this however does not change the fit regions much. We do not include in the fit the data from the ATIC-2 and 4~\cite{ATIC-2, ATIC-4} and PPB-BETS~\cite{PPB-BETS} balloon experiments, that in particular had found a peak at about 500-800 GeV (possibly indicative of a DM mass around the TeV scale), as this feature has later been questioned by the \FERMI\ findings. Since however \FERMI+\HESS\ also pin down the mass in a similar range, our results would easily apply also to the case in which ATIC is reincluded.} with the use of DM generated $e^+,e^++e^-$ and $\bar p$ spectra, as discussed in detail in~\cite{CKRS}: we find the best fit values by scanning over the propagation parameters of charged cosmic rays and over the uncertainties on the slope and normalization of the astrophysical electron, positron and antiproton background. Our analysis is again conservative in the fact that we consider the slopes and normalizations of the backgrounds of the different species as uncorrelated, while a significant correlation exists (e.g. both positrons and $\bar p$ are produced in the same collisions as astrophysical secondaries). A proper treatment of this effect would reduce the ``allowed parameter space" for DM models.
We do not include any galactic boost factor due to substructures within the Milky Way halo. We also stress again that we consider only the case in which the annihilation or decay proceeds directly into a SM particle-antiparticle pair. I.e. we do not address here the cases of one-step annihilation models~\cite{onestep}, of decay into three leptons (see e.g.~\cite{decay3}) etc.

The resulting allowed regions on the plane $m_\chi$--$\langle \sigma v \rangle$ are shown in fig.s~\ref{fig:exclusionann} and~\ref{fig:exclusionannH} for the annihilating DM case and on the plane $m_\chi$--$\tau_{\rm dec}$ in fig.~\ref{fig:exclusiondecay} for the decaying DM one. 
The regions that allow to fit the \PAMELA\ data alone (for positrons and antiprotons) are individuated by green and yellow bands, for 95\% C.L. and 99.999\% C.L., corresponding to $\Delta \chi^2(m_\chi,\langle \sigma v\rangle) \equiv \chi^2(m_\chi,\langle \sigma v\rangle) -\chi^2_{\rm min} \simeq 6$ and $\Delta \chi^2 \simeq 23$, with 2 d.o.f. (analogously for decay). 
The addition of the \FERMI\ and \HESS\ data allows to individuate a narrow range for the mass of the DM particle, around a few TeV. It also selects a DM that annihilates or decays into leptons only, since, for such a mass, the excess in antiprotons that would arise from non-leptonic channels is excluded by \PAMELA\ $\bar p$ data. The regions that fit \PAMELA+\FERMI+\HESS\ combined are represented in the figures by red and orange areas (for the same confidence levels as above). Notice that the smoothness of the \FERMI\ spectrum forbids a reasonable fit with the ${\rm DM}\, {\rm DM} \to e^+e^-$ or ${\rm DM} \to e^+e^-$ channel, that would produce too peaked a feature: the reduced $(\chi^2_{\rm min})_{\rm red}$ turns out to be well above 2 for all DM profiles, so we plot no allowed region.
The typical annihilation cross sections that are required are of order $10^{-23}\, {\rm cm}^3/{\rm sec}$ up to $10^{-20} \,{\rm cm}^3/{\rm sec}$ or more, depending on the mass of the candidate and the annihilation channel. For decay, $\tau_{\rm dec} \approx 10^{26}$ sec is needed.

\section{Results}\label{results}
\begin{figure}[t]
\begin{center}
\hspace{-8mm}
\includegraphics[width=0.333\textwidth]{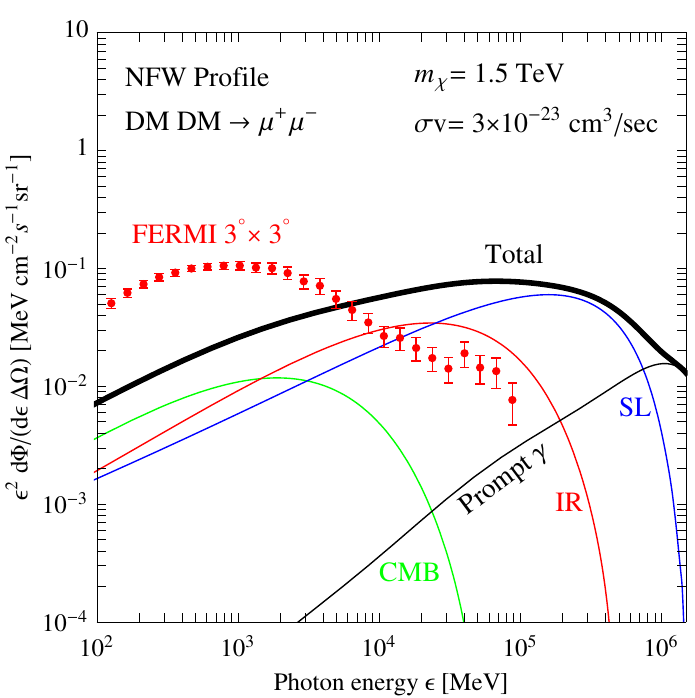}\
\includegraphics[width=0.333\textwidth]{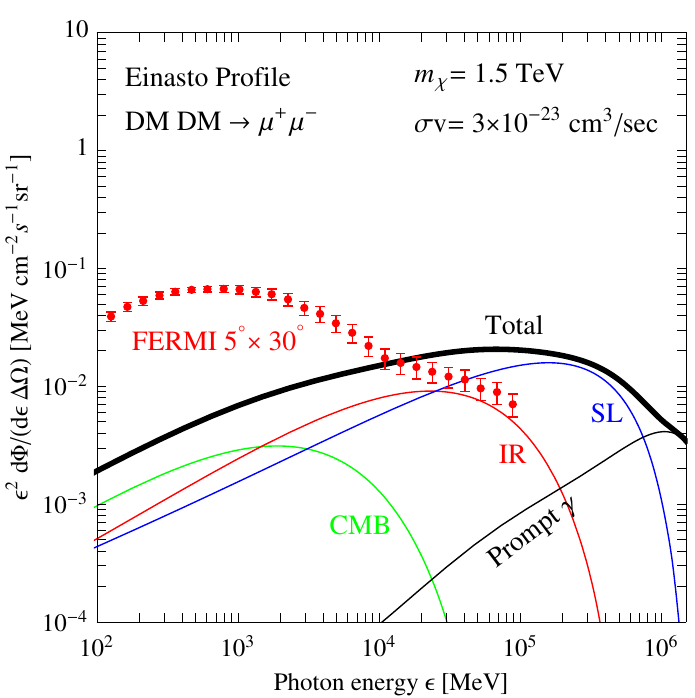}\
\includegraphics[width=0.333\textwidth]{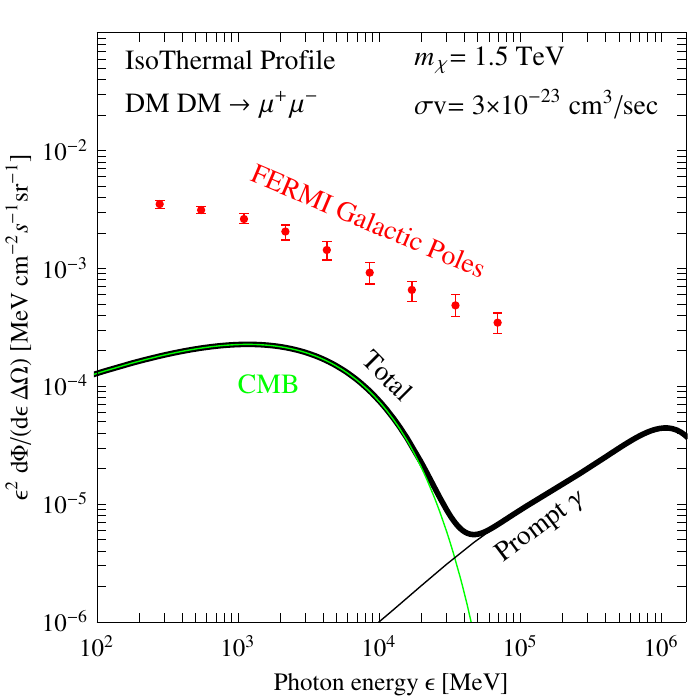}\\[2mm]
\hspace{-8mm}
\includegraphics[width=0.333\textwidth]{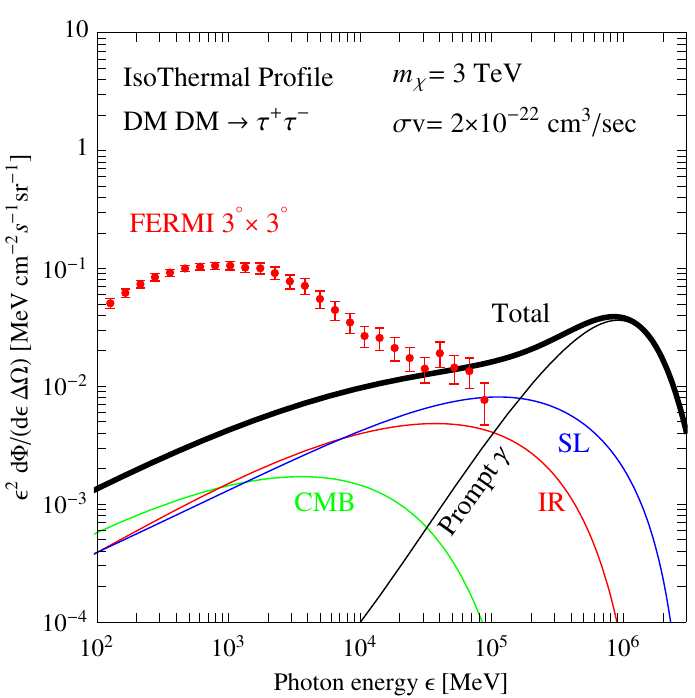}\
\includegraphics[width=0.333\textwidth]{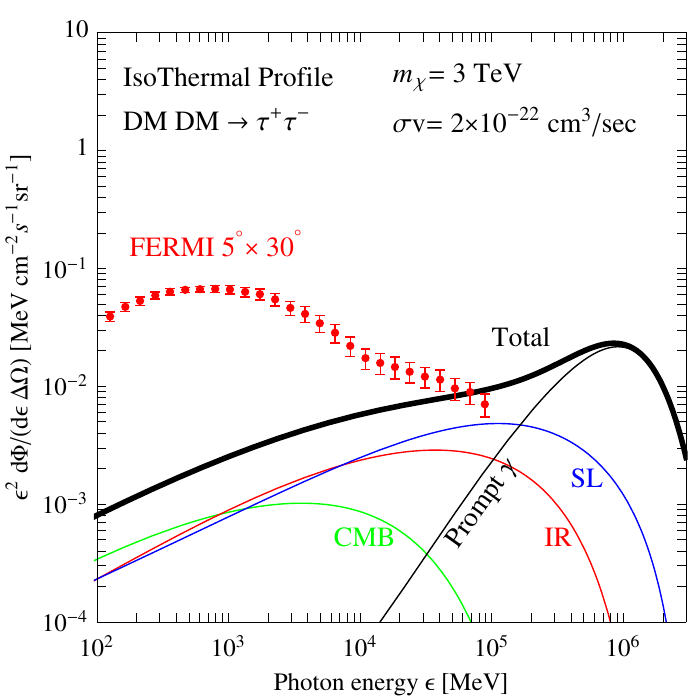}\
\includegraphics[width=0.333\textwidth]{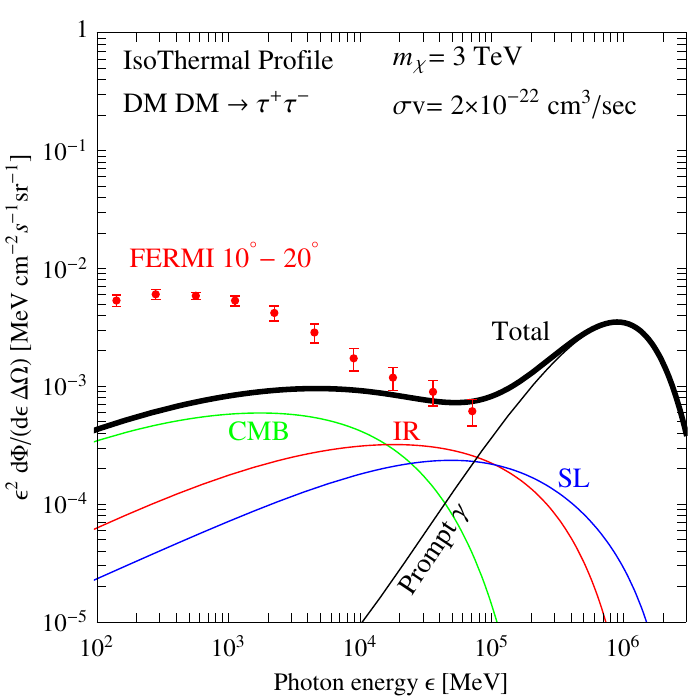}
\caption{\em\label{fig:fluxes} 
Gamma ray fluxes for a few sample DM candidates, compared to the \FERMI\ datapoints in the different observation regions that we consider. See text for details.}
\end{center}
\end{figure}

We start by reporting, in Fig.~\ref{fig:fluxes}, the gamma-ray fluxes for a few typical TeV DM candidates with large annihilation cross section in leptonic channels (of the type invoked to explain the anomalies in $e^\pm$ data), in different angular windows and for different halo profiles. For each case we plot the total gamma-ray flux and its different components: the prompt gamma-ray emission and the ICS emission on StarLight (SL), on InfraRed light (IR) and on the CMB. 
Similar plots can be drawn for the decaying DM case. 

\medskip

As apparent, in all these cases the spectral shapes of the curves of the DM signals are very different from what is observed.
In a $\epsilon^2\,\d \Phi/\d \epsilon$ plot, the \FERMI\ data point show a decreasing behaviour (that often seems well-accounted
for by a simple, likely astrophysical, power-law), while a curve rising up to $\sim\,$TeV energy, possibly with a ``double bump'' feature (characteristic of the high energy prompt and low energy ICS emissions) is expected from DM.  This immediately reasserts that a significant astrophysical signal is needed to account for the data, confirming the conservative approach of our analysis.

The first two panels of Fig.~\ref{fig:fluxes} show the predicted signal in the `$3^\circ \times 3^\circ$' and `$5^\circ \times 30^\circ$' regions from a DM candidate of mass 1.5 TeV, annihilating with 100\% B.R. into $\mu^+\mu^-$ with a cross section of 3 10$^{-23}$ cm$^3$/sec, assuming an NFW or  Einasto (i.e. those suggested by numerical N-body simulations) respectively. It is evident that the predicted signal overshoots the data points, very evidently in the first case but also significantly in the second case. These kind of DM candidates are therefore clearly excluded by observations.

The third panel of Fig.~\ref{fig:fluxes} shows the predicted signal in the `Galactic Poles' region for the same DM candidate, assuming an Isothermal profile. In this case the signal lies well below the \FERMI\ data points. This kind of scenario cannot be excluded or explored yet in this observational region.

The second row of panels in Fig.~\ref{fig:fluxes} shows the current situation for the Isothermal cored profile choosing a 3 TeV DM annihilating into $\tau^+\tau^-$ with a cross section of 2 10$^{-22}$ cm$^3$/sec. The \FERMI\ data points in {\it each} region are in clear tension with the predicted DM signal. Note that for the purposes of the following plots we shall not depict these models as ``firmly excluded'', although this would clearly be the conclusion of a proper likelihood analysis {\it combining} the different
channels (not to speak of further constraints from non-$\gamma$ channels).

As an aside, the plots in Fig.~\ref{fig:fluxes} also allow to appreciate the different spectral features of the predicted DM signals originating from the different ISRF in the different regions. For the inner regions (`$3^\circ \times 3^\circ$' and `$5^\circ \times 30^\circ$') the ICS signal is predicted to be dominated by the contribution on StarLight. For the intermediate region (`$10^\circ-20^\circ$ strips') the lower energy contribution on CMB is more important. At the `Galactic Poles' the CMB contribution only is considered: the other components are subdominant. The prompt contribution is particularly relevant for the $\tau^+\tau^-$ channel, as it originates from the $\pi^0 \to \gamma \gamma$ from the hadronic $\tau^\pm$ decays. At the higher latitudes it is well distinct from the IC emission, as the latter is located at lower energies.

\bigskip

The above examples guide us in the interpretation of the exclusion plots, discussed below.  
We derive the constraints on the annihilation cross section or the decay half-life by the following conservative prescription. For each observational region with the corresponding data points, fixed an annihilation channel, a DM distribution profile and a DM mass, we impose that the ICS signal must not exceed any of the experimental data points by more than 3$\sigma$. This determines a maximum annihilation cross section or a minimum half life. Notice once again that we do not assume anything on the astrophysical gamma ray background, although it must be clearly present and dominant by visual inspection of the plots. We will return on this point in Sec.~\ref{discussion}.

\begin{figure}[!htp]
\begin{center}
\hspace{-8mm}
\includegraphics[width=0.333\textwidth]{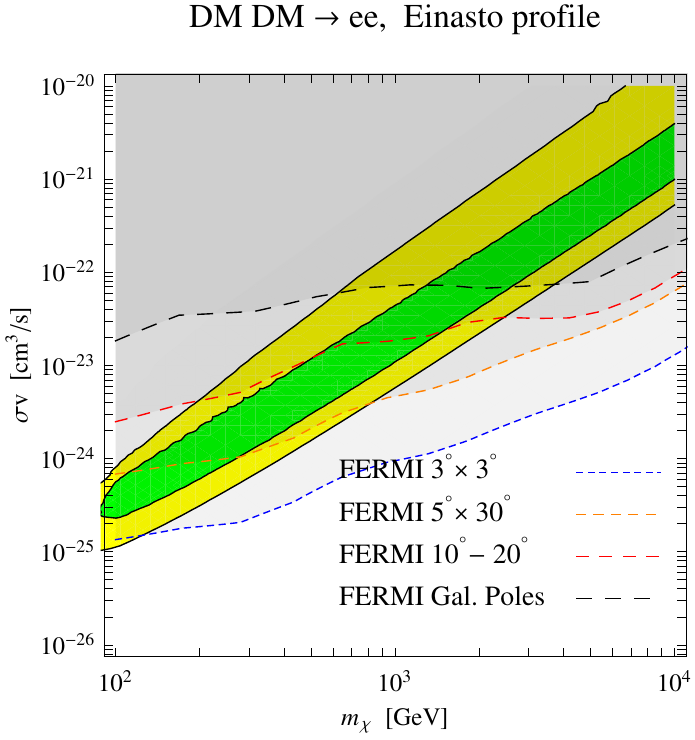}\
\includegraphics[width=0.333\textwidth]{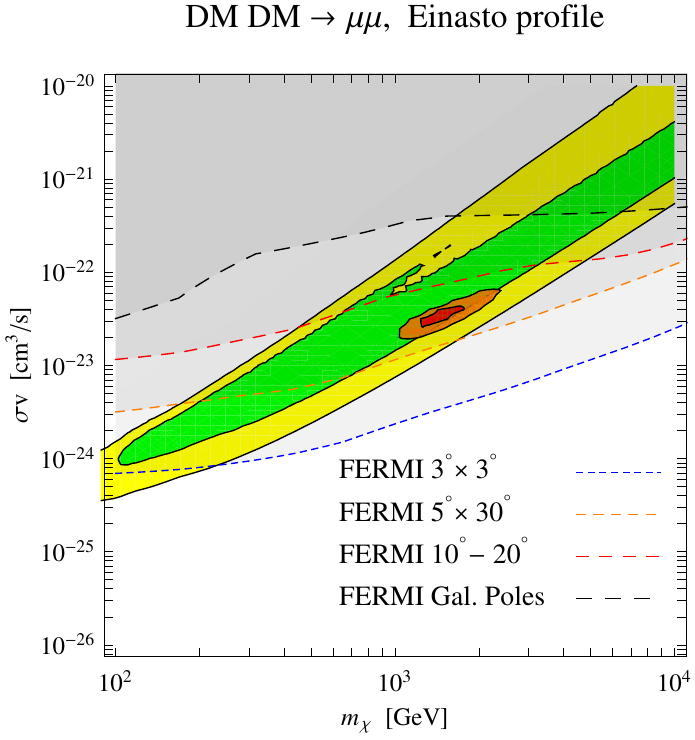}\
\includegraphics[width=0.333\textwidth]{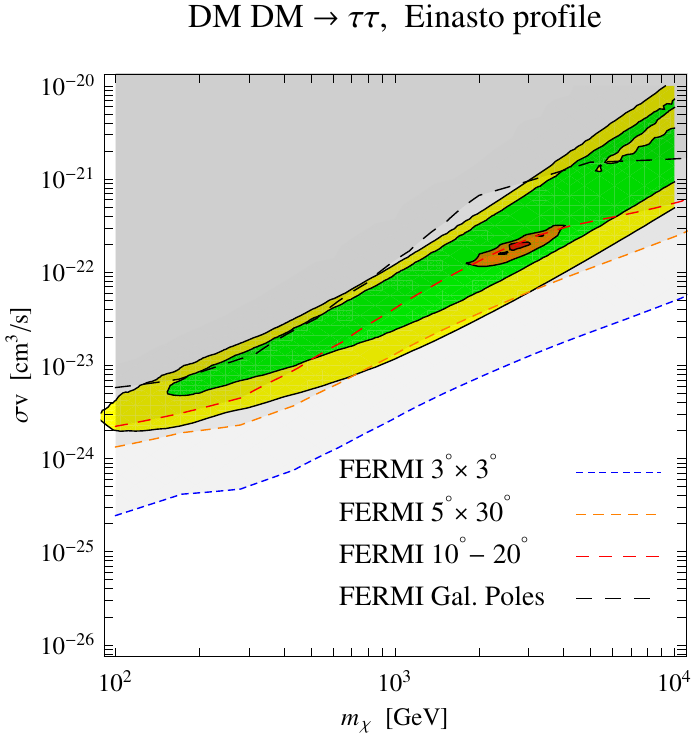}\\[2mm]
\hspace{-8mm}
\includegraphics[width=0.333\textwidth]{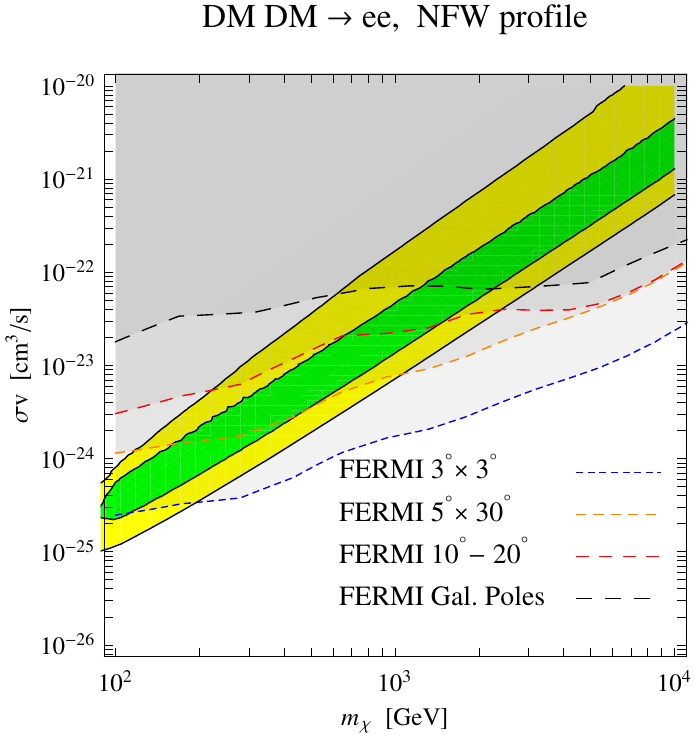}\
\includegraphics[width=0.333\textwidth]{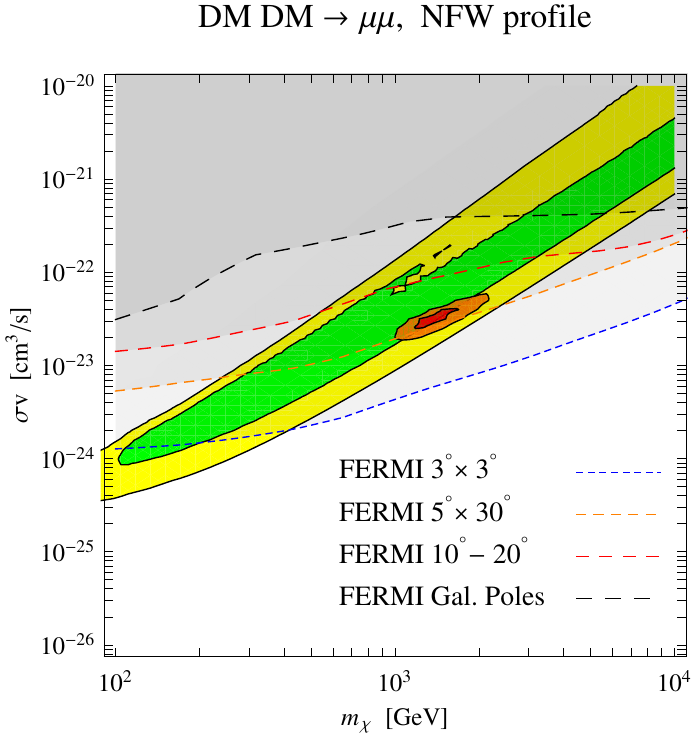}\
\includegraphics[width=0.333\textwidth]{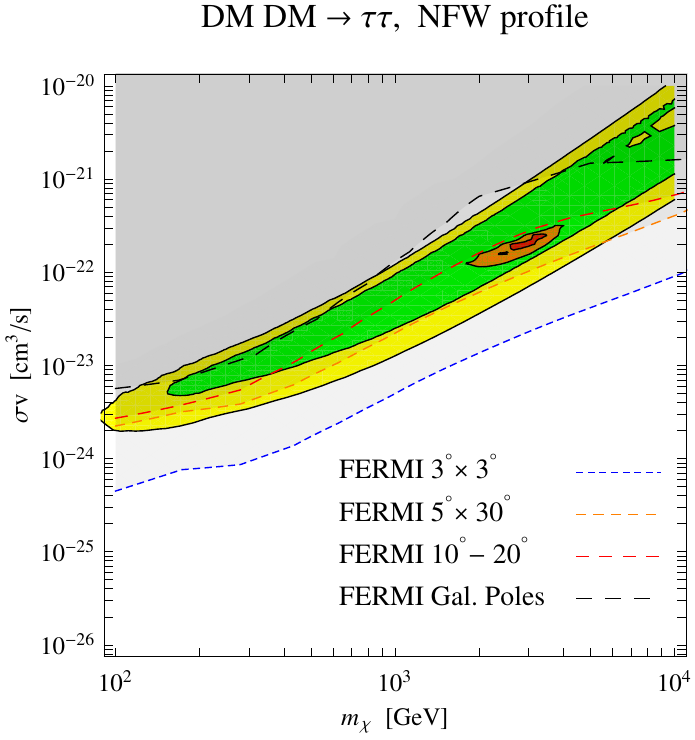}\\[2mm]
\hspace{-8mm}
\includegraphics[width=0.333\textwidth]{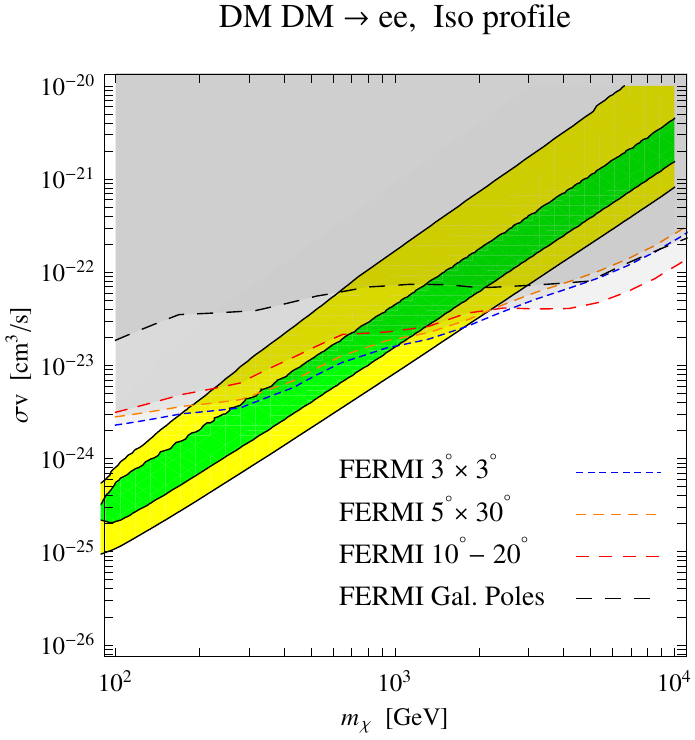}\
\includegraphics[width=0.333\textwidth]{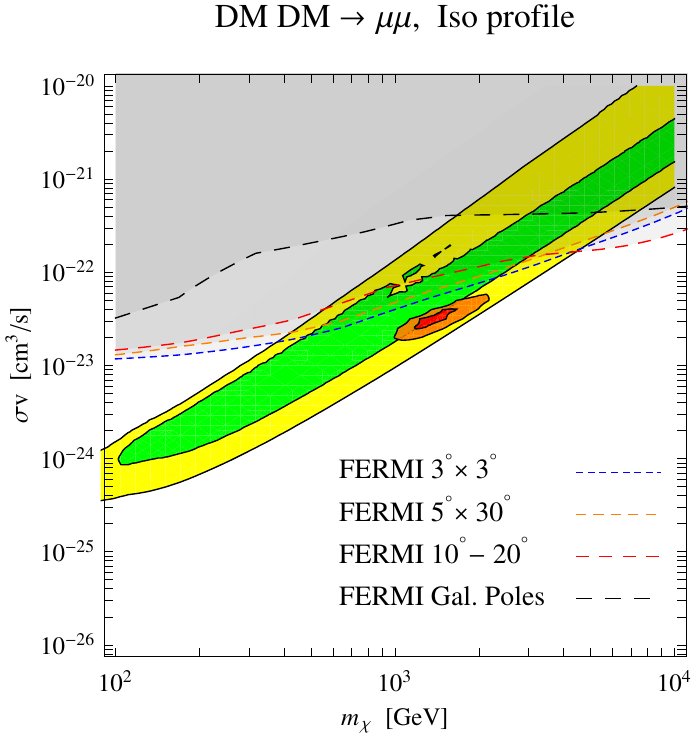}\
\includegraphics[width=0.333\textwidth]{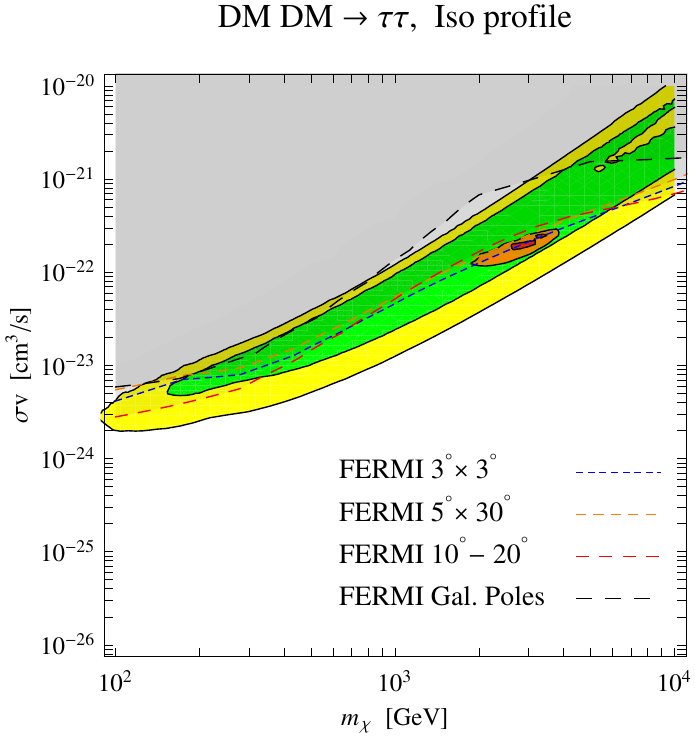}
\caption{\em\label{fig:exclusionann} 
The regions on the parameter space $m_\chi$--$\sigv$ that are excluded by the diffuse galactic gamma ray measurements by the \FERMI\ satellite. The first column of panels refers to DM annihilations into $e^+e^-$, the second into $\mu^+\mu^-$ and the third into $\tau^+\tau^-$; the three rows assume respectively an NFW, an Einasto and a cored Isothermal profile. Each panel shows the exclusion contour due to \FERMI\ observations of the `$3^\circ \times 3^\circ$' region (blue short dashed line), `$5^\circ \times 30^\circ$' region (orange dashed line), the `$10^\circ - 20^\circ$ strip' (red long dashed line) and the `Galactic Poles' $|b| > 60^\circ$ region (black long dashed line). We also report the regions that allow to fit the \PAMELA\ positron data (green and yellow bands, 95 \%  and 99.999 \% C.L. regions) and the \PAMELA\ positron + \FERMI\ and \HESS\ data (red and orange blobs, 95\%  and 99.999\% C.L. regions).}
\end{center}
\end{figure}

\begin{figure}[!ht]
\begin{center}
\hspace{-8mm}
\includegraphics[width=0.333\textwidth]{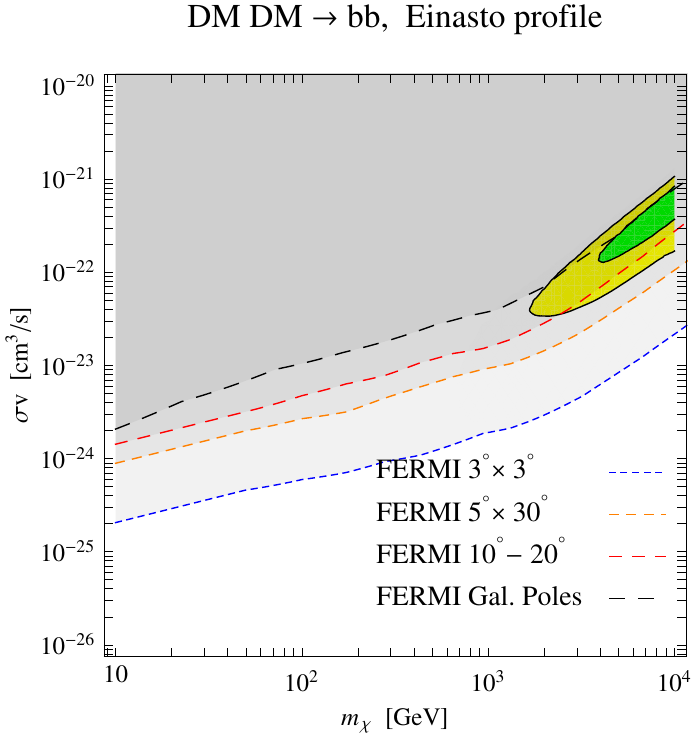}\
\includegraphics[width=0.333\textwidth]{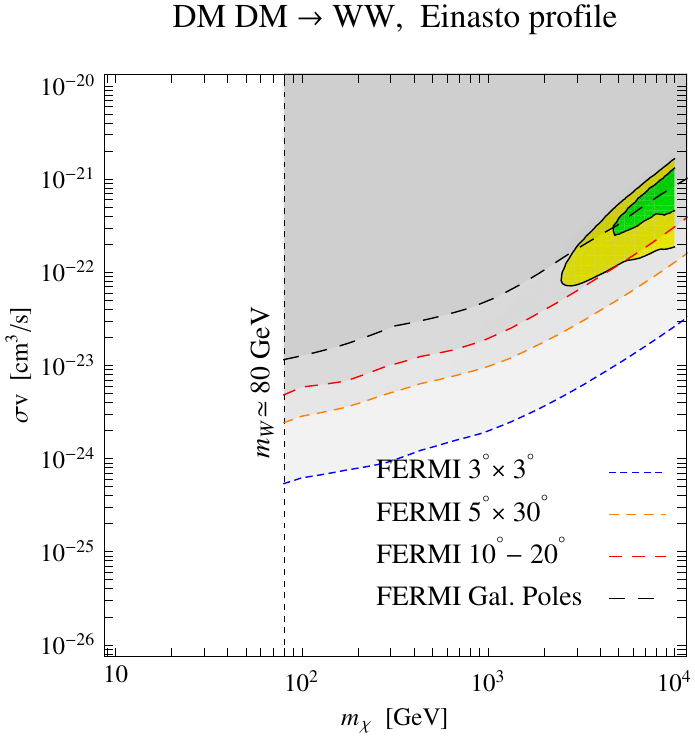}\
\includegraphics[width=0.333\textwidth]{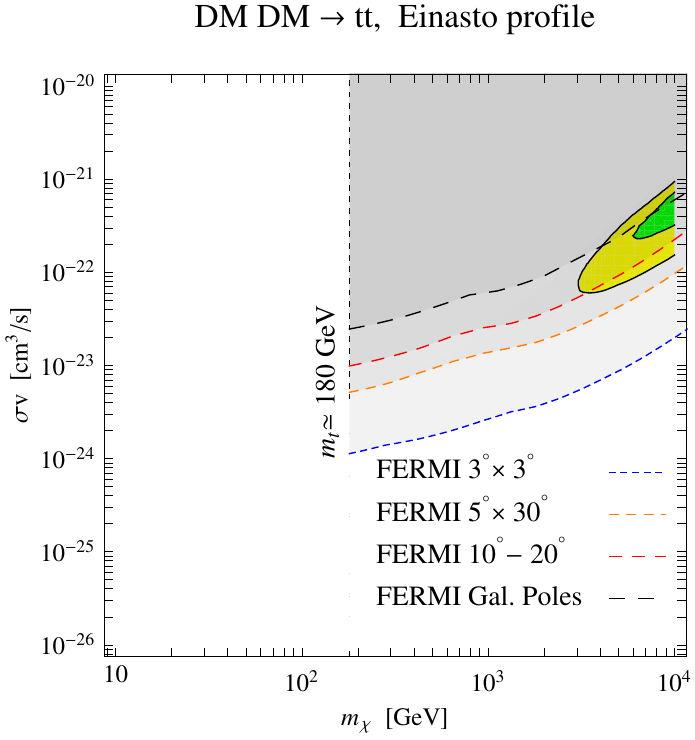}\\[2mm]
\hspace{-8mm}
\includegraphics[width=0.333\textwidth]{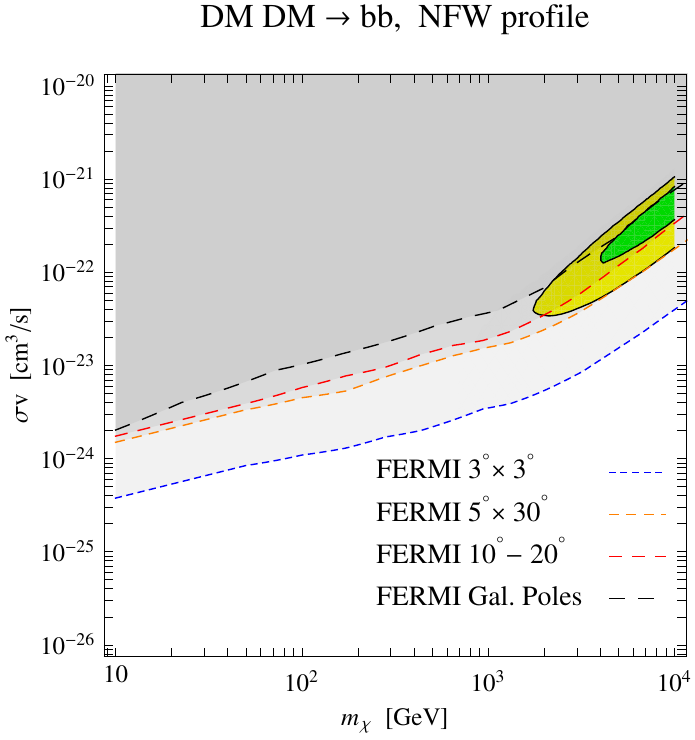}\
\includegraphics[width=0.333\textwidth]{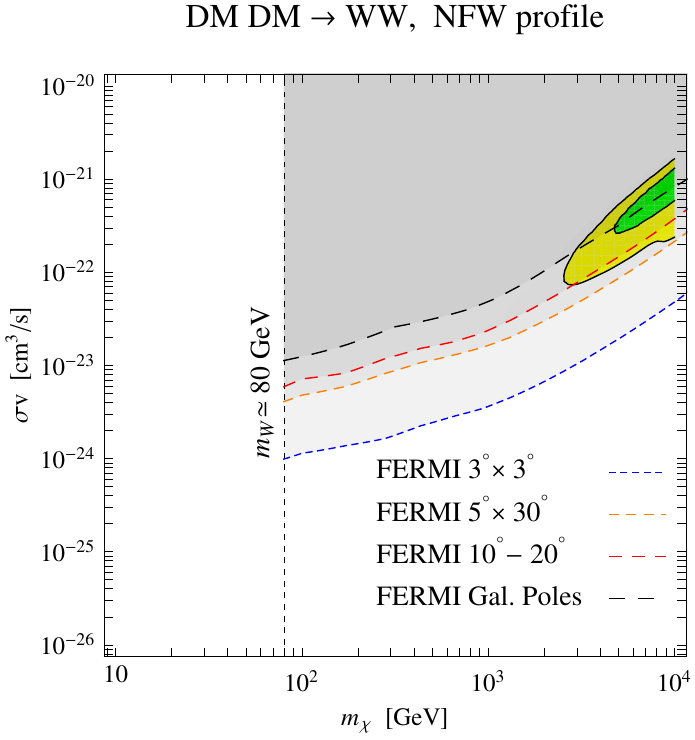}\
\includegraphics[width=0.333\textwidth]{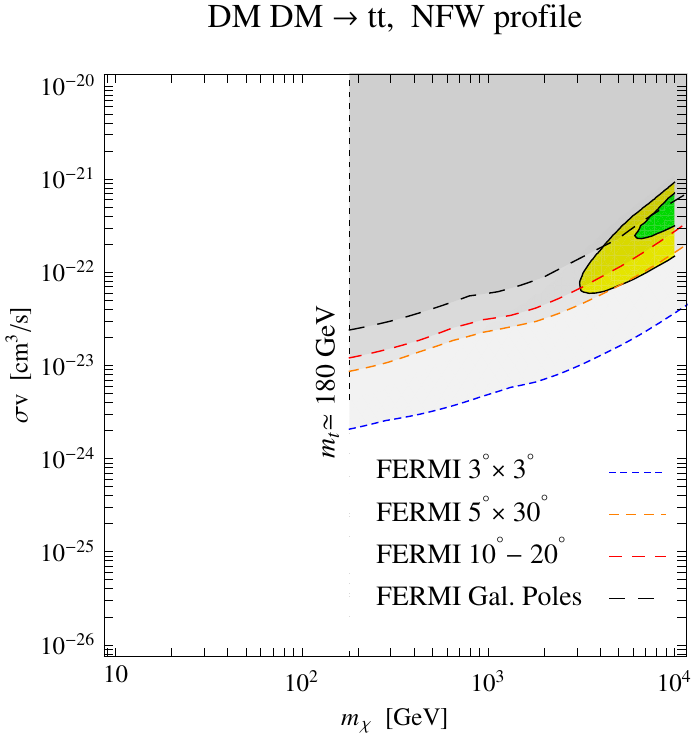}\\[2mm]
\hspace{-8mm}
\includegraphics[width=0.333\textwidth]{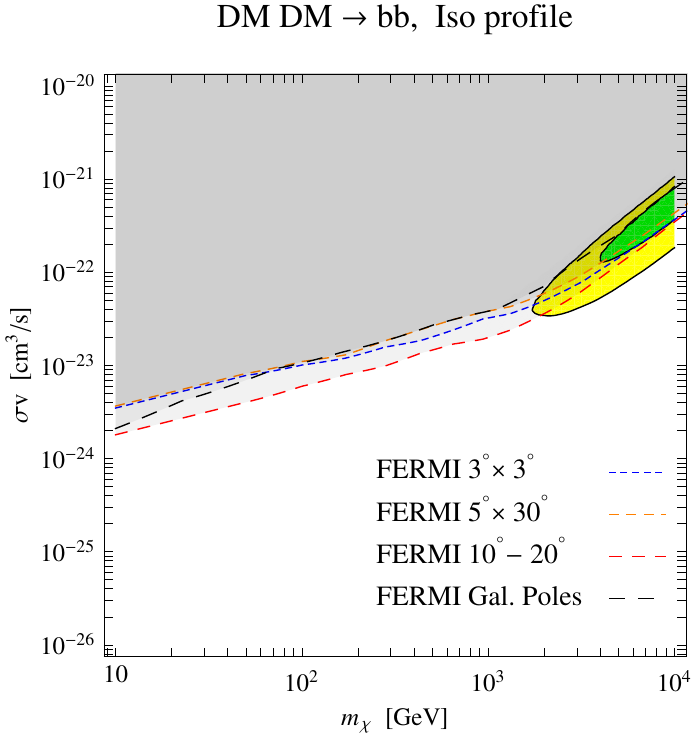}\
\includegraphics[width=0.333\textwidth]{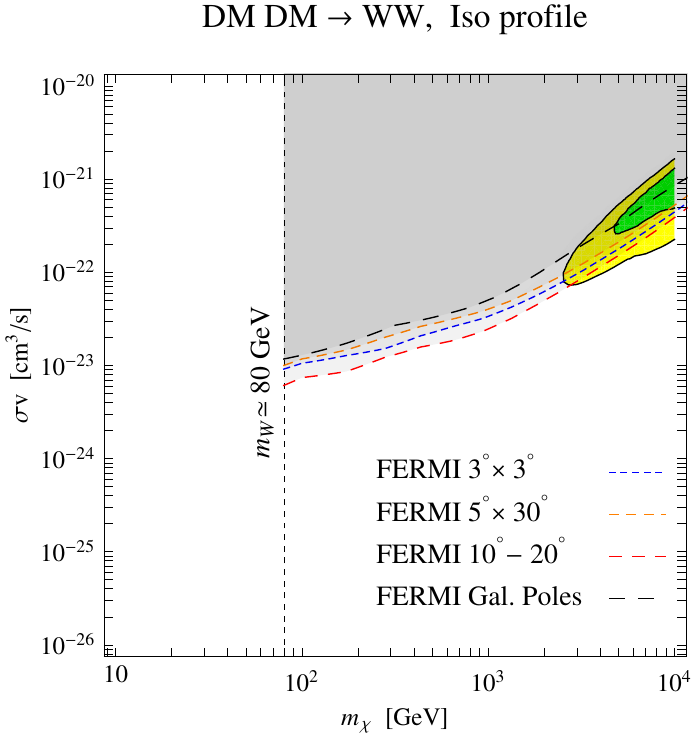}\
\includegraphics[width=0.333\textwidth]{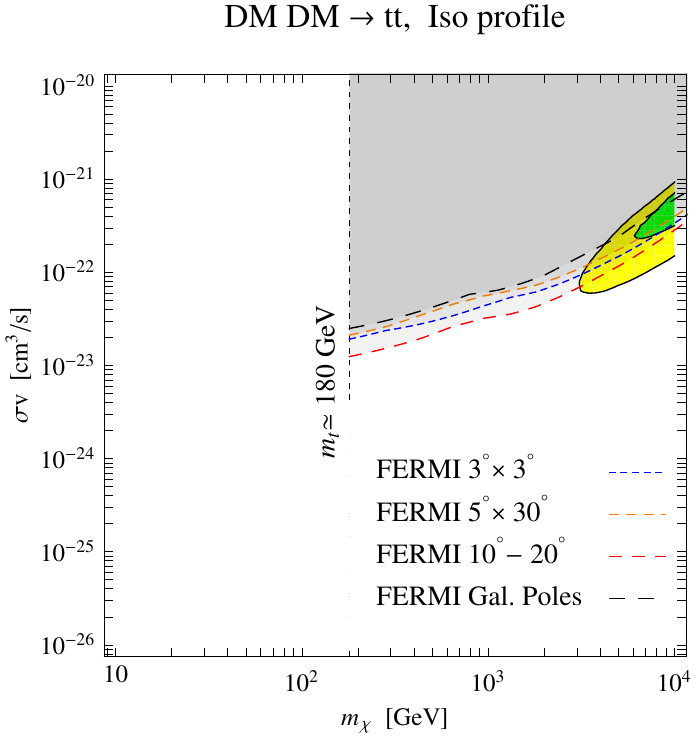}
\caption{\em\label{fig:exclusionannH} 
Like figure\fig{exclusionann}, but for DM annihilations into $b\bar{b}$, $W^+W^-$,  and $t\bar{t}$.}
\end{center}
\end{figure}

\begin{figure}[tp]
\begin{center}
\hspace{-8mm}
\includegraphics[width=0.333\textwidth]{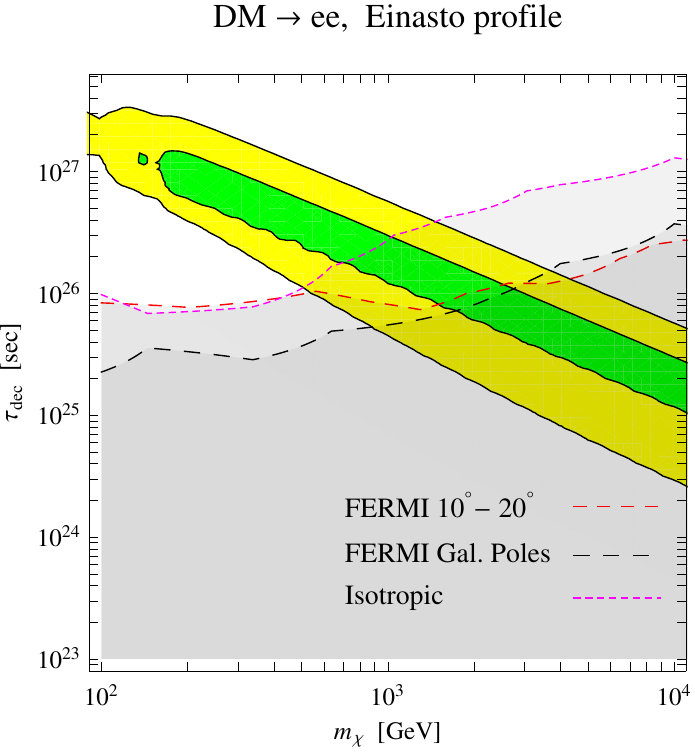}\
\includegraphics[width=0.333\textwidth]{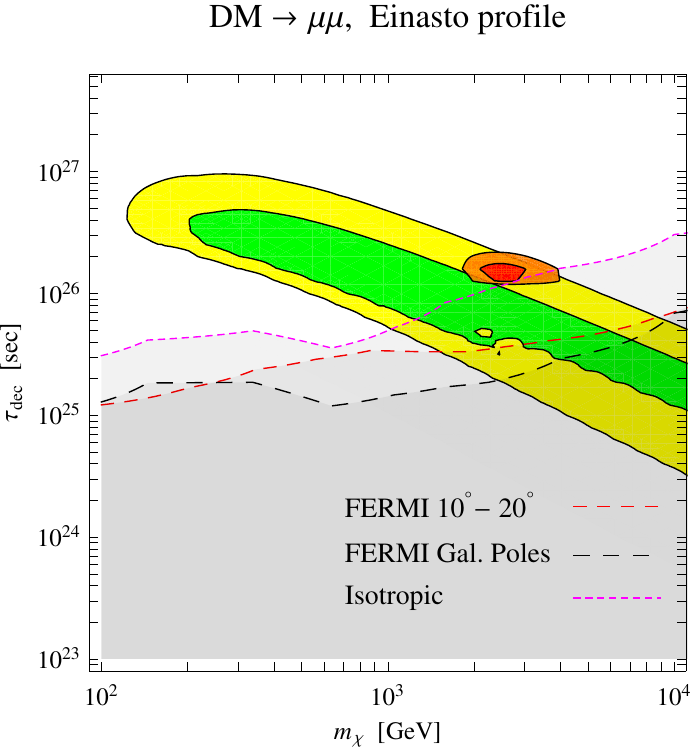}\
\includegraphics[width=0.333\textwidth]{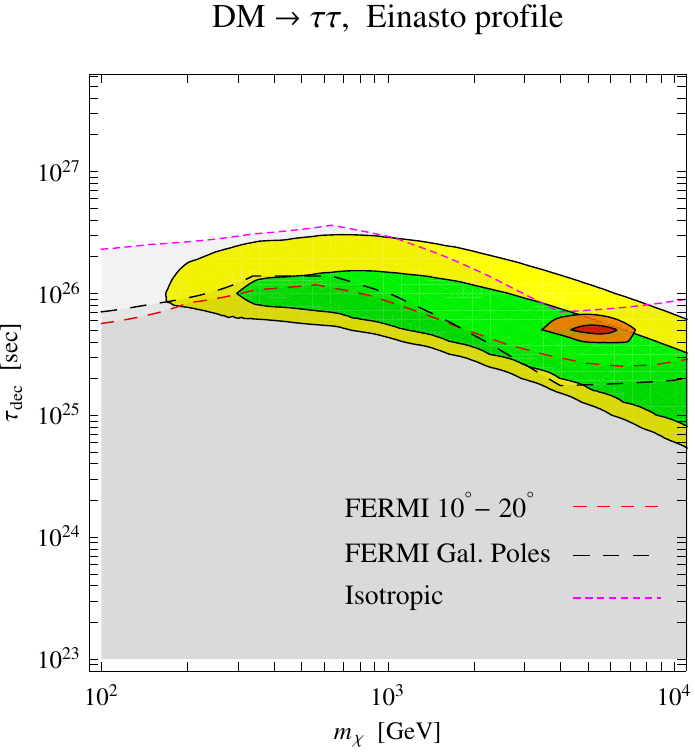}
\hspace{-8mm}
\caption{\em\label{fig:exclusiondecay} 
Similarly to figure\fig{exclusionann} but for decaying Dark Matter. The vertical axis reports here the half-life $\tau_{\rm dec}$ in seconds. The exclusion contours are due to \FERMI\ observations of the `$10^\circ - 20^\circ$ strip' (red dashed line), the $|b| > 60^\circ$ `Galactic Poles' region (black long dashed line) and the isotropic flux (magenta dotted line). We also report the regions that allow to fit the \PAMELA\ positron data (green and yellow bands, 95\%  and 99.999\% C.L. regions) and the \PAMELA\ positron + \FERMI\ and \HESS\ data (red and orange blobs, 95\%  and 99.999\% C.L. regions) in terms of decaying Dark Matter. We here report only the case of an Einasto galactic DM profile: the cases of an Isothermal or a NFW profile are essentially identical (see text for details).}
\end{center}
\end{figure}

\noindent  In Fig.~\ref{fig:exclusionann}, we report the exclusion plots for leptonic annihilation modes. We find that:
\begin{itemize}
\item[$\circ$] For profiles of the NFW or Einasto type, i.e. those suggested by numerical N-body simulations,
 current data from the inner Galaxy exclude not only DM scenarios explaining simultaneously \FERMI+\HESS\ and \PAMELA\ features, but also \PAMELA\ data alone to a high confidence level. The only small region left in parameter space (and with a low confidence level) is the one at  $m_\chi$ just above 100 GeV, i.e. just above the highest energy point presented by \PAMELA. This implies a final test  at AMS-02 of the DM origin of the \PAMELA\ anomalous positron fraction: should the ``excess'' continue to rise well above 100 GeV, the corresponding DM mass would enter the region already excluded, in clear conflict with the \FERMI\ diffuse gamma ray data analyzed here.
\item[$\circ$] For ``cored'' profiles, the data fail to fully exclude, {\it in this simplified analysis}, the regions explaining \PAMELA+\FERMI+\HESS\ features. The $\tau^{\pm}$ channel is however already in strong tension with existing data (and likely the $2\div 3\sigma$ best fit region would be excluded to significant  confidence level by a combined analysis of the $\gamma$-channels), while the $\mu^{\pm}$ annihilation is still allowed by a factor $\sim 2\div 3$.
\item[$\circ$]  For NFW or Einasto profiles, already this extremely conservative analysis gives constraints on $\sigv$ which, for typical masses $m_\chi\simeq 100\,$GeV, are above the value expected for a S-wave annihilating thermal relic by a factor of order 10 or less, especially for the $e^{\pm}$ mode.
\end{itemize}

\noindent In Fig.~\ref{fig:exclusionannH}, we report the exclusion plots for hadronic annihilation modes.  We can see that:
\begin{itemize}
\item[$\circ$] In the ranges where the different channels are open, the bounds are almost independent from the channel: this follows from the quasi-universality of the gamma spectrum arising from the fragmentation of sufficiently ``heavy'' SM particles.
\item[$\circ$] For NFW or Einasto profiles, a significant contribution to the positron fraction appears strongly excluded by gamma-rays. Note that for cored-profiles, the inner-galaxy data are relaxed, but the higher latitude ones are not and those are sufficient to suggest an (at most) subleading contribution of 
DM to the positron flux.
\item[$\circ$] For NFW or Einasto profiles, already this extremely conservative analysis gives constraints on $\sigv$ which, for typical masses $m_\chi\simeq 100\,$GeV, are within one order of magnitude of the value expected for a S-wave annihilating thermal relic. 
\end{itemize}

\noindent  Finally, in Fig.~\ref{fig:exclusiondecay} we report the exclusion plot for the case of leptonically decaying DM. 
We show that:
\begin{itemize}
\item[$\circ$] The intermediate and high latitude galactic observations impose significant constraints, that are in general stronger that any other Galactic gamma ray or neutrino constraint (see e.g.~\cite{MPSV}). They are however insufficient to probe the \PAMELA+\FERMI+\HESS\ regions. 
\item[$\circ$] The residual isotropic radiation measured by \FERMI\ imposes the strongest constraint. It excludes the decay explanation for the \PAMELA+\FERMI+\HESS\ anomalies for the $\tau^+\tau^-$ channel {\it independently} on the profile. Even the allowed region for the $\mu^+\mu^-$ channel starts to be constrained by the data and it is reasonable to expect that a more refined analysis with
future data will be able to probe definitely this region as well.
\item[$\circ$] Note that in Fig.~\ref{fig:exclusiondecay} we report only the case of the Einasto DM profile: the exclusions plots for the NFW or cored isothermal cases are essentially identical. This is to be expected: the allowed \PAMELA+\FERMI+\HESS\ regions bear a very weak dependence on the profile choice, as positrons and electrons come from the local halo region where all profiles resemble one another. The DM galactic gamma ray signal from intermediate and high latitudes (and therefore the constraints) vary only within 5\% or less, as illustrated by the values of the $\bar J_{\rm dec}$ factors in Table~\ref{tab:regions} (the isotropic flux is obviously halo independent as already discussed).
\end{itemize}

\section{Discussion and Conclusions}\label{discussion}

In this article, we have provided a first assessment of the power that new data on the diffuse emission from the Fermi Gamma-ray Space Telescope have in constraining Dark Matter indirect signals. Even under the very brutal approximation of neglecting any astrophysical background contributing to the signal and using conservatively 3 $\sigma$ exclusion criteria, current data from the inner Galaxy (e.g. `$3^\circ \times 3^\circ$') exclude a benchmark DM mass $m_\chi\simeq 100\,$GeV if its annihilation is larger than  a factor 5$\div 30$ (depending on the channel) of the typical $\sigv \simeq 3\times10^{-26}\,$cm$^3$/s, when profiles suggested by N-body simulations are employed. Higher-latitude constraints are a factor $\sim 10$ weaker and comparable to constraints for cored profiles. It is remarkable that already such a simplified analysis is powerful enough to explore regions of parameter space not excluded otherwise, providing constraints which are comparable to or better than those obtained by the Fermi collaboration by
analyzing dwarf spheroidals~\cite{Abdo:2010ex} or the isotropic signal from cosmological DM annihilation~\cite{Abdo:2010dk}.
This confirms, if needed, the Galactic halo as the ``target of excellence'' for constraining or detecting gamma rays from DM.

\begin{figure}[t]
\begin{center}
\hspace{-8mm}
\includegraphics[width=0.333\textwidth]{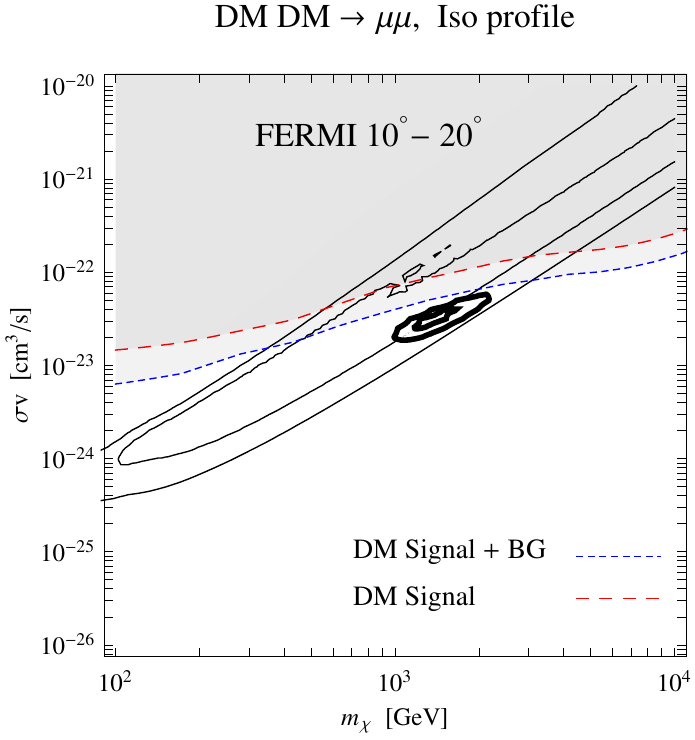}\ \ \ \ \
\includegraphics[width=0.333\textwidth]{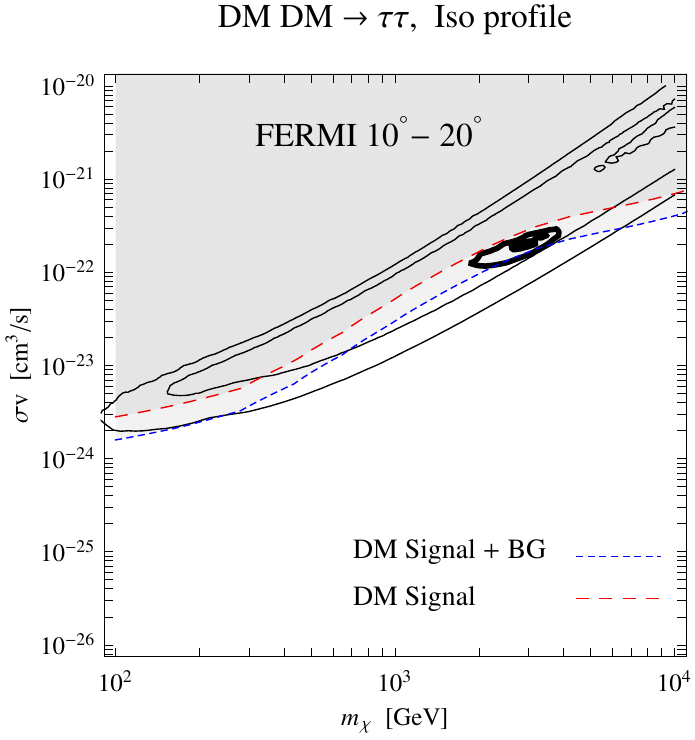}\
\caption{\em\label{fig:withBG} 
For one selected observational region (the `$10^\circ-20^\circ$ strips') and profile, and for two exemplar annihilating DM candidates, we show the current bounds derived by our conservative procedure of requiring that the DM signal must not exceed the observational data points by more than 3$\sigma$ (red dashed line) and the prospective regions that can be explored by \FERMI\ if an astrophysical gamma ray background is assumed and the exclusion criterion is relaxed to 2$\sigma$ (blue short dashed line).}
\end{center}
\end{figure}

On the other hand, the absence of astrophysical background is an extremely (unrealistically) conservative assumption as visual inspection of the plots in Fig.~\ref{fig:fluxes} confirms. In the pre-\FERMI\ era,  some studies have been performed showing the possible improvement  in sensitivity when accounting for pointlike and diffuse sources in the Galactic Center region (see e.g.~\cite{Dodelson:2007gd}). The current high-quality data certainly allow one to improve over these exploratory studies to forecast the ultimate \FERMI\ sensitivity to DM. While a proper treatment of this problem goes beyond our current purposes, in Fig.\fig{withBG} we present for illustration the exclusion plots that would follow from the {\it current} `$10^\circ-20^\circ$ strips' data if its bulk could be robustly attributed to astrophysical processes, as in the adjusted propagation model shown in~\cite{Porter} and the exclusion criterion is relaxed from $3\sigma$ to $2\sigma$. The `improvement' is about a factor of 2. Likely, intermediate-latitude DM bounds could be made competitive with current conservative inner-galaxy constraints. In turn, the latter could  improve significantly if maps were  cleaned from further astrophysical sources contaminating the total flux: notice that the `$3^\circ \times 3^\circ$' degree field data {\it are not} corrected for pointlike sources~\cite{Digel}, which have already been identified by the very \FERMI\ satellite~(see e.g.~\cite{Cohen-Tanugi}).
It appears realistic to expect that by the final stages of the \FERMI\ mission a better knowledge of the astrophysical sources (both pointlike and diffuse) would allow for significantly stronger constraints or perhaps enough sensitivity for a detection of a ``benchmark DM candidate'', in particular if the DM halo profile is  steep enough toward the GC, or if significant substructures are present enhancing the diffuse flux. This would be essentially consistent with early expectations~\cite{Baltz:2008wd}. 

\bigskip

We have also addressed the issue of the consistency of DM interpretation of the leptonic CR spectra 
from \PAMELA, \FERMI, \HESS\ with the new diffuse gamma ray data. It is worth stressing that, even before such new \FERMI\ data, the annihilating DM interpretation of the leptonic CR spectra was challenged by the constraints from gamma rays and radio waves from the galactic center~\cite{BCST,MPV,PPB}, from the integrated cosmological flux of ICS photons~\cite{Profumo,HooperICS,Raidal}, from neutrinos from the galactic center~\cite{neutrinos,MPSV} and from Big Bang Nucleosynthesis~\cite{BBN}.  Among the most robust constraints are those from CMB observations and the reionization history of the Universe~\cite{Galli,Slatyer,CIP,Raidal,Kanzaki}, which rule out (if barely) the best fit regions of  \PAMELA+\FERMI+\HESS.

The new diffuse gamma ray data allow a further test of the DM hypothesis, in particular of the prompt and especially IC emission at relatively large latitudes (for former studies with previous data see e.g.~\cite{Borriello:2009fa,CP,MPSV}; for related studies with current data see~\cite{Ibarraanisotropy,PS}). These signals have a weaker dependence  on the extrapolations of the profile in the inner region of the Galaxy than most previously cited constraints. For example,  the strip $10^\circ - 20^\circ$ does not depend at all on the profile in the inner $\sim 1.5\,$kpc of the Galaxy and most of the signal comes actually from a significantly larger distance, due to geometric effects.  Moreover the new data allow for to stringently probe the {\em decaying} DM explanations of the anomalies.

\medskip

In summary, here are our results: 
\begin{itemize}
\item Decaying DM appears significantly excluded from the isotropic emission inferred from the \FERMI\ data, unless it decays mostly/exclusively in $\mu^{\pm}$.
\item We find that DM annihilating hadronically --already excluded by antiproton data as a joint explanation for the $e^+$ and $e^++e^-$ anomalies-- cannot even account for a significant fraction of the positron fraction. This remains essentially true even for cored DM profiles. 
\item DM annihilating leptonically is excluded as leading explanation for CR charged lepton data in all cases if the halo shape is close to the Einasto or NFW one, and disfavoured even for cored profiles for $\tau^{\pm}$ mode. It appears that only for cored halo profiles, mostly annihilating into $\mu^{\pm}$, some viable fit of the CR leptons can be found. However, it is reasonable that even such ``fined-tuned'' situations may run into some troubles when scrutinized more deeply. 
\end{itemize}
These conclusions are in general based on robust and conservative procedures, as discussed in the text. We list here a few fine points and caveats. 

 In realistic models where DM annihilates into several channels --including quarks ($\to \pi$'s) and gauge bosons-- the B.R. into $\mu^{\pm}$ might be insufficient to evade the bounds. A similar effect would derive from considering three-body channels with final state electroweak radiation, which in particular for these engineered ``leptophilic'' scenarios could have non-negligible B.R.~\cite{Kachelriess:2009zy}. 
 
  Also, when moving from the attitude of constraining an otherwise undetected DM to the  hypothesis of {\it checking predictions} of a DM candidate postulated to account for some other CR signal, omitting astrophysical background in the gamma-channel is not a self-consistent approach. Indeed, in order to account for $e^{+}$ and $e^{-}$ data, some additional background
must be introduced and some choices for propagation parameters have been made (e.g. normalization and index of
diffusion function, equivalent to a ``grammage'' in  a leaky box picture). The low-energy $e^+$ fraction is explained by invoking CR spallation and $\pi\to\mu\to e$ production, etc.
Thus, any ``best-fit choice'' with background plus DM is accompanied by a corresponding {\it lower limit}  for the diffuse gamma-ray background due to CR spallation in the ISM. While in principle other sources of gamma's not showing-up in electrons are possible, the one above is unavoidable. It is worth reminding that predictions based on the ``conventional'' model calculated by GALPROP~\cite{Galprop} and recently refined by the \FERMI\ team on the light of their new data~\cite{Porter} fit the data quite satisfactorily without the need for a leading role of DM.   A similar caveat applies to the practice of best-fitting {\it independently} the background for different channels: since different astrophysical backgrounds are correlated, a proper analysis may restrict the parameter space for DM models to fit the data.  

\medskip

We conclude this article by a few comments on how to modify/improve the constraints. 
\begin{itemize}
\item[$\triangleright$] The \HESS\ ($e^++e^-$) data points might contain a sizable contamination from diffuse isotropic gamma rays, up to 50\%~\cite{HESSleptons}. Strictly speaking, for a given DM candidate it is therefore the sum of the ($e^++e^-$) yield and the isotropic $\gamma$ (as computed in this paper) which has to be compared to the \HESS\ data. However, for the typical candidates that we considered, the isotropic gamma ray flux turns out to be much smaller than the corresponding ($e^++e^-$) flux. No more stringent bounds can therefore be derived by this procedure. For candidates with larger mass (above several TeV) this constraint could however turn out to be significant.
\item[$\triangleright$] As already mentioned, should the \FERMI\ collaboration present data or limits on the flux at the highest energies (e.g. the $E \sim 300\,$GeV bin) they would be very powerful in constraining DM models that still provide marginal fits to the data, provided that the associated error bars prove to be small enough. Even an upper limit at the level of the current detected flux around $\sim 100\,$GeV would strengthen the bounds by a factor $\sim$2 or so. 
\item[$\triangleright$] Also, we have not scanned the \FERMI\ maps looking ``a posteriori'' for the most constraining regions for a given DM model starting directly from the $\gamma-$maps. If anything, such strategy would strengthen the constraints.
\item[$\triangleright$] Moreover, note that \FERMI\ has presented some early results on the limits on gamma-ray line-emission~\cite{Murgia}. This emission is unavoidable at loop level, but more model dependent. For a typical B.R.$\sim 10^{-3}$ and for relatively light candidates (up to a 200 (400) GeV for annihilating (decaying) DM) the resulting bounds should be comparable or stronger than the ones reported in the text. 
\item[$\triangleright$] Additional constraints are expected from the associated neutrino production (unavoidable already from the two-body channel for the $\mu^{\pm}$ and $\tau^{\pm}$ modes). 
It is likely that a reanalysis of the Super-K data, including large regions around the Galactic Center, may provide a conclusive test of these models (or, alternatively, an analysis of future Icecube+DeepCore data~\cite{Mandal}.)
\end{itemize}

\paragraph{Acknowledgements.} We thank Alessandro Strumia for useful discussions and {\em constant attention} to our work. We thank Simona Murgia and Alejandro Ibarra for very useful discussions. The work of P.P. is supported in part by the International Doctorate on AstroParticle Physics (IDAPP) program. We thank the EU Marie Curie Research \& Training network ``UniverseNet" (MRTN-CT-2006-035863) for support.

\bigskip
\appendix

\section{{\large Uncertainties of the constraints due to galactic magnetic field models and $e^\pm$ synchrotron radiation}}
In this appendix, we address the issue of how the constraints that we presented are affected by the uncertainties on the galactic magnetic field. 
Indeed, in our analysis we have always considered (in line with the analysis of~\cite{CP}) the energy losses suffered by the DM-produced electron and positrons into synchrotron radiation as subdominant. If the magnetic field is however significantly large, then competing synchrotron energy losses can be important, the IC emission can be reduced and the propagation of $e^\pm$ fitting the \PAMELA+\FERMI+\HESS\ modified.\footnote{We stress that of course only the IC compton emission from DM is affected by the value assumed for the magnetic field $B$. This is the dominant signal only for the leptonic modes, in particular $e^+e^-,\mu^+\mu^-$. The prompt component (which is e.g. present or dominant in the $\tau^+\tau^-$ annihilation or decay channel) is not affected.}

For $E_e\gg\,$GeV and in the Thomson regime, the energy loss scales roughly as (e.g.~\cite{Schlick},
Eq.~(4.6.1))
\begin{equation}
\frac{\d E}{\d t}\propto - E^2 [B_\perp^2+27\,u_\gamma]
\end{equation}
where $B_\perp$ is the perpendicular component of the magnetic field with respect to the direction of motion, measured in $\mu$G, and $ u_\gamma$ is the energy density in low-energy photons, measured in eV.
In the vicinity of the solar system, $u_\gamma^\odot\approx 1$ in these units (see e.g.~\cite{Schlick},
Table.~(2.2)), while in the GC region one has $u_\gamma^{\rm GC}\approx 10$ (see for example 
Fig.1 in~\cite{CP}). From the relation above, we can derive that as long as 
$B_\perp^\odot\ll 5.2$ and $B_\perp^{\rm GC}\ll 16.4$, the IC losses dominate. Equivalently, for an isotropic field for which $B_\perp\approx \sqrt{2/3}B_{tot}$,  the IC dominates as long as
\begin{equation}
\{B_{tot}^\odot< 6.4,\:B_{tot}^{\rm GC}< 20.1\}, \quad {\rm (in \ } \mu {\rm G} {\rm \ units)}. \label{boundsB} 
 \end{equation}
On the other hand, the interest of DM annihilation/decay final states like $e^+e^-,\mu^+\mu^-$ is mostly related to phenomenological reasons, namely
the fit of  \FERMI\ and/or \PAMELA\ lepton data. So, rather than the absolute level of the bounds, it is more interesting to know how the relative position of the best-fit region vs. bounds moves with changing parameters. 
In the (physically unlikely, see below) limit where the magnetic loss term dominates over IC, what alters this position of, say the GC bounds from IC compared to the fit, are (to a first approximation) differences in the ratio $(B_\perp^\odot)^2/u_\gamma^\odot$ (the quantity relevant to the local propagation of the $e^\pm$ that fit \PAMELA+\FERMI+\HESS) compared to $(B_\perp^{\rm there})^2/u_\gamma^{\rm there}$ where with the notation $^{\rm there}$ we want to indicate that this is intended evaluated in the regions associated with the window of gamma ray observation and therefore it is the quantity relevant to energy losses suffered by the $e^\pm$ that produce the bound. For the large regions in our analysis, the conditions  $^{\rm there}$ are very much expected to be similar to the conditions here, i.e. at $^\odot$.
A larger sensitivity to the above considerations is instead obtained when considering the bounds coming from the $3^\circ\times 3^\circ$ region around the GC. We consider already the bounds coming from this region as being the least robust, due to the fact the we neglect the effect of diffusion. 

\medskip

In order to estimate the possible error introduced by the ignorance on the $B$-field, first of all we consider the models discussed in~\cite{Kachelriess:2005qm}. We perform an average of $B^2$ over a kpc scale volume, whose square root we dub  for simplicity of notation $\langle B^{\odot,{\rm GC}}\rangle$.
This is the typical propagation scale of high-energy electrons and roughly the size of the $3^\circ\times 3^\circ$ region. For the two models (TT and HMR) which are regular towards the GC region, one gets values well within the requirements of Eq.~(\ref{boundsB}), for example \{$\langle B^{{\rm GC}}\rangle,\langle B^{\odot}\rangle\}=\{4.6, 1.4 \}$ for the TT model.\\ 
These models however only aim at describing the large scale features of the field as inferred e.g. from Faraday Rotation measurements. Since additional power may be present in the smaller scale field to which synchrotron radiation is responding, too, a perhaps more appropriate representative for the total field can be taken from~\cite{Strong:1998fr},
\begin{equation}
B_{tot}=B_0\exp(-(r-R_{\odot})/R_B)\exp(-|z|/z_B)\,.
\end{equation}
The suggested parameters $R_B$=10 kpc, $B_0$=6.1 $\mu$G, and  $z_B=$ 2 kpc  reproduce  the absolute magnitude and profiles of the 408 MHz radio emission; for these values  one gets $\{\langle B^{{\rm GC}}\rangle,\langle B^{\odot}\rangle\}=\{5.4,\,12\}$, again consistent with the requirements of Eq.~(\ref{boundsB}). 

Finally, one can consider models which extrapolate to a very high value towards the GC, like the 
 PS one in~\cite{Kachelriess:2005qm}, where there is a mathematical divergence at the GC position. Those kind of models  have to be regularized based on physical arguments. In~\cite{LaRosa:2005ai} it was extensively argued that the value of the {\it pervasive} magnetic field in the inner degrees of the Galaxy is likely $\sim 10\,\mu$G, i.e. well below the critical value of $\sim 20\,\mu$G reported above
 and consistent with the estimates from previous fitting formulae. On the other hand, high fields of the order $\sim 100\,\mu$G or larger might be present in relatively thin filaments (see~\cite{Crocker:2010gy} for references).  For the purposes of this paper, it is sufficient to report as a general argument that as long as 
 \begin{equation}
\left(\frac{B_{fil}}{20\,\mu{\rm G}}\right)^2\times \frac{V_{fil}}{1{\rm kpc}^3}\ll 1\,
\end{equation}
the limits in Eq.~(\ref{boundsB}) are satisfied and the bounds derived in the main text are robust. For example, if filaments were to occupy a (100 pc)$^3$ volume, where the field reaches  $\sim 100\,\mu$G
field, the above limit is easily fulfilled. What if even the above condition is violated? Of course, the IC bound {\it from the inner Galaxy} would be weakened, since most of the energy loss would be damped into synchrotron radiation. The recent analysis in~\cite{Crocker:2010gy} however showed that, when combining radio and gamma-data, the overall constraints on the annihilation cross section are quite robust, independently of the $B$-field assumed, although {\it which channel} yields the best constraint has a $B$-dependence. 

\medskip

At the light of these results, we conclude that the IC bounds are quite robust against variations in $B$ field, except possibly from the region very close to the GC and for extreme values of parameters. In the latter case, however, deriving more reliable constraints from IC is not only model-dependent, but actually of little relevance, since other bounds are known to be as stringent (or more stringent) than the IC ones.


\footnotesize
\begin{multicols}{2}
  
\end{multicols}

\end{document}